\documentclass[superscriptaddress,preprint,showpacs,preprintnumbers,amsmath,amssymb,nofootinbib,toc,12pt]{revtex4}
\pdfoutput=1
\usepackage{latexsym}
\usepackage{graphicx}
\usepackage{amsthm}
\usepackage{float}
\usepackage{amsmath}
\usepackage{graphicx,subfigure}
\usepackage{fancybox,feynmp}
\usepackage{indentfirst}
\usepackage{epsfig}
\usepackage{epsfig,subfigure,psfrag}
\usepackage{dcolumn}
\usepackage{bm}
\usepackage{slashed}
\usepackage{cancel}
\usepackage{color}
\usepackage{tabularx}
\usepackage{multirow}
\usepackage{hyperref}
\hypersetup{
    colorlinks=true,       
    linkcolor=blue,          
    citecolor=blue,        
    filecolor=blue,      
    urlcolor=blue,           
    linktoc=page
}

\preprint{LA-UR-17-29295}

\newcommand{\be}{\begin{eqnarray}}
\newcommand{\ee}{\end{eqnarray}}
\newcommand{\beq}{\begin{eqnarray}}
\newcommand{\eeq}{\end{eqnarray}}

\newcommand{\bea}{\begin{eqnarray}}
\newcommand{\eea}{\end{eqnarray}}
\newcommand{\eV}{{\rm eV}}
\newcommand{\keV}{{\rm keV}}
\newcommand{\MeV}{{\rm MeV}}
\newcommand{\GeV}{{\rm GeV}}
\newcommand{\TeV}{{\rm TeV}}
\newcommand{\Br}{\text{Br}}
\newcommand{\lsim}{\lesssim}
\newcommand{\gsim}{\gtrsim}

\newcommand{\Qpq}{Q^{_\text{PQ}}}
\newcommand{\Tr}{\text{Tr}}
\newcommand{\s}{\text{s}}
\newcommand{\U}{\text{U}}
\newcommand{\SU}{\text{SU}}

\newcommand{\da}[1]{{\bf \color{red} D.A.: #1}}

\begin{document}

\title{A viable QCD axion in the MeV mass range}

\author{Daniele S. M. Alves}
\email{spier@lanl.gov}
\affiliation{Center for Cosmology and Particle Physics, Department of Physics, New York University, New York, NY 10003}
\affiliation{Department of Physics, Princeton University, Princeton, NJ 08544}
\affiliation{Theoretical Division, Los Alamos National Laboratory, Los Alamos, NM 87545, USA}

\author{Neal Weiner}
\email{neal.weiner@nyu.edu}
\affiliation{Center for Cosmology and Particle Physics, Department of Physics, New York University, New York, NY 10003}

\date{\today \\ \vspace*{1cm}}

\begin{abstract}

The QCD axion is one of the most compelling solutions of the strong CP problem. There are major current efforts into searching for an ultralight, {\it invisible} axion, which is believed to be the only phenomenologically viable realization of the QCD axion. {\it Visible} axions with decay constants at or below the electroweak scale are believed to have been long excluded by laboratory searches. Considering the significance of the axion solution to the strong CP problem, we revisit experimental constraints on QCD axions in the $\mathcal{O}$(10 MeV) mass window. In particular, we find a variant axion model that remains compatible with existing constraints. This model predicts new states at the GeV scale coupled hadronically, and a variety of low-energy axion signatures, such as rare meson decays, nuclear de-excitations via axion emission, and production in $e^+e^-$ annihilation and fixed target experiments. This reopens the possibility of solving the strong CP problem at the GeV scale.

\end{abstract}

\maketitle

\tableofcontents

\newpage

\numberwithin{equation}{section}
\renewcommand\theequation{\arabic{section}.\arabic{equation}}
\section{\label{sec:intro} Introduction}

The Peccei-Quinn (PQ) mechanism is arguably the most compelling solution of the strong CP problem. Soon after it was originally proposed \cite{Peccei:1977hh, Peccei:1977ur}, it was realized that a light pseudoscalar would emerge in the infrared spectrum as a manifestation of the underlying PQ mechanism - the QCD axion \cite{Weinberg:1977ma, Wilczek:1977pj}. As the parameters characterizing the QCD axion, such as mass and decay constant, span several orders of magnitude, its phenomenology changes dramatically, with implications ranging from hadronic physics to astrophysics and cosmology.
For the past four decades, this rich phenomenology has been explored over an ever broadening range of decay constants, $\mathcal{O}(\GeV)\lsim f_a\lsim \mathcal{O}(M_{\text{Pl}})$ \cite{Kim:1986ax, Cheng:1987gp, Kim:2008hd, Hagmann:2008zz, Marsh:2015xka, Patrignani:2016xqp}, and there is still an ongoing and vigorous experimental effort to test the QCD axion.

The existing consensus is that {\it visible} QCD axions, {i.e.,} those with decay constants at or below the weak scale, have long been excluded by laboratory searches \cite{Peccei:1988ci, Turner:1989vc}, such as beam dump experiments, rare meson decays, and nuclear de-excitations.\footnote{For $f_a \gsim 100~\text{GeV}$,
the QCD axion is also constrained by stellar evolution \cite{Raffelt:1990yz, Engel:1990zd, Raffelt:2006cw}, and most recently by a combination of the CMB power spectrum and primordial $^4$He and D/H abundances \cite{Millea:2015qra}.} This has motivated the formulation of {\it invisible} axion models \cite{Kim:1979if, Shifman:1979if, Dine:1981rt, Zhitnitsky:1980tq}, which, combined with further astrophysical bounds from stellar evolution, CMB and BBN, redirected experimental efforts to extremely weakly coupled axions ($f_a\gsim 10^{9}~\GeV$) \cite{Sikivie:1985yu, Peng:2000hd, Mueller:2009wt, Asztalos:2009yp, Bahre:2013ywa, Aalseth:2002qf, Irastorza:2011gs, Armengaud:2014gea, Budker:2013hfa, Kahn:2016aff, TheMADMAXWorkingGroup:2016hpc, Graham:2015ouw}, which are ultra-light ($m_a\lsim 10^{-3}~\eV$), and could be non-thermal dark matter \cite{Preskill:1982cy,Abbott:1982af,Dine:1982ah}.

Given the significance of the QCD axion and the many resources dedicated to probing its existence, it is important to ensure that no caveats have been overlooked, and that no gaps in already probed regions of parameter space were left uncovered. Motivated by this, we revisit constraints on the MeV mass window for the QCD axion and its variants, and discuss a particular realization of the QCD axion that has not yet been definitively excluded. Our results reopen the possibility that the strong CP problem might be solved below the weak scale, and suggest new, hadronically-coupled degrees of freedom at the GeV scale.

In Secs.\,\ref{sec:genericAxions}-\ref{kaons} of this paper, we refute previous, premature conclusions that the MeV mass range for the QCD axion has been completely ruled out. We then discuss a viable axion variant and its couplings to photons, nucleons, and electrons, and the relevant experimental implications (Secs.\,\ref{GammaGamma}-\ref{other}). Finally, we comment on UV completions of such variants and associated phenomenology (Secs.\,\ref{GeVcompletion}-\ref{EWcompletion}). For an outline of this paper, vide Table of Contents.

\section{\label{sec:genericAxions} Constraints on Generic MeV Axions}

Generic constraints on QCD axions in the MeV mass window can be broadly classified into three categories: (i) amenable to model-building, (ii) plagued by large hadronic uncertainties, and (iii) evaded only by {\it pion-phobia}.  The first category of constraints can be evaded by well-established model-building tools. Constraints in the second category are more difficult to avoid with model-building, but suffer from significant uncertainties which preclude them from fully and unambiguously ruling out the MeV mass range for the QCD axion. Finally, the third category encompasses the strongest constraints, which can only be avoided by a special class of axion variants which are {\it pion-phobic}, i.e., which have suppressed mixing with the neutral pion. While this can be achieved by model-building to some degree, the extreme {\it pion-phobia} needed to avoid exclusion also depends critically on the light quark mass ratio being close to a ratio of Peccei-Quinn charges (this will become clear in Sec.\,\ref{dolor}). Indeed the most up-to-date determinations of $m_u/m_d$ indicate that it is very close $1/2$, making extreme {\it pion-phobia} a realistic possibility.

\subsection{Constraints Amenable to Model-Building}
\label{avoidable}

We start by discussing the main experimental observables that have excluded generic QCD axions in the MeV mass window. Despite being severe, these constraints can be ``model-built away'' by deviating from generic models.

If the axion couples to heavy quark flavors, such as charm or bottom, it is strongly constrained by radiative decays of quarkonia, such as $J/\psi\rightarrow\gamma a$ and $\Upsilon\rightarrow\gamma a$. Wilczek \cite{Wilczek:1977zn} showed that such decay widths can be related to the leptonic widths via:
\beq
\Gamma(J/\psi\rightarrow\gamma a) ~&=&~ \Gamma(J/\psi\rightarrow\mu^+\mu^-)\; \frac{\lambda^2_{c}}{2\pi\alpha} \;C_{J/\psi},\\
\Gamma(\Upsilon\rightarrow\gamma a) ~&=&~ \Gamma(\Upsilon\rightarrow\mu^+\mu^-)\; \frac{\lambda^2_{b}}{2\pi\alpha} \;C_{\Upsilon},
\eeq
where $\lambda_{c}$ and $\lambda_{b}$ are the axion couplings to $\bar c\, i\gamma_5 c$ and $\bar b\, i\gamma_5 b$, respectively, and $C_{J/\psi}$, $C_{\Upsilon}\sim\mathcal{O}(1)$ encode QCD and relativistic corrections \cite{Nason:1986tr}.
The MeV mass range for the QCD axion corresponds to decay constants $f_a$ in the $\mathcal{O}(1-10)$ GeV range, and consequently to large couplings of the axion to charm and/or bottom quarks, namely, $\lambda_{c}\sim\mathcal{O}(m_c/f_a)$ and/or $\lambda_{b}\sim\mathcal{O}(m_b/f_a)$. With such large couplings, radiative decays of quarkonia to $\gamma a$ would dominate over leptonic modes, in gross contradiction with observation. In fact, bounds from quarkonia decays alone \cite{Albrecht:1986ht,Hsueh:1992ex,Armstrong:1992wu,Bai:1995ik,Adams:2005mp} were sufficient to exclude the entire parameter space of the original PQWW axion (see, for instance, \cite{Davier:1986ps}). These and other bounds led to much activity during the 1980's in model building QCD axion {\it variants}, i.e., variations of the original PQWW axion that could evade existing constraints at the time and remain viable. 

A class of visible axion variants that trivially evades quarkonia bounds are those that couple the axion exclusively to first generation quarks.\footnote{At least at tree level --- axion couplings two second and third generations, including flavor non-diagonal couplings, will invariably be generated radiatively upon electroweak and PQ symmetry breakings. These couplings however are sufficiently suppressed to avoid present bounds, and will be ignored for the remainder of this paper.} Nevertheless, they still have to contend with other constraints. For instance, if such axions have suppressed couplings to leptons and decay dominantly to a pair of photons, they are sufficiently long lived and hence robustly excluded by a variety of beam dump experiments \cite{Bergsma:1985qz,Bjorken:1988as,Davier:1989wz,Blumlein:1990ay} in the range $100~\keV\lsim m_a\lsim 30~\MeV$. Limits from beam dumps are substantially degraded if axions couple to electrons with strength $\mathcal{O}(m_e/f_a)$. In this case, axions heavier than 5 -- 10 MeV become very shortly lived, $ \tau_a\lsim 10^{-13}~\s$, and their decay products stop in the earth shielding before reaching the detector, evading this class of bounds. Short-lived axions decaying to $e^+e^-$ also evade severe constraints from
$K^+\rightarrow\pi^+(a\rightarrow\text{invisible})$, whose branching ratio is bounded to be
$\lsim 4.5\times10^{-11}$ \cite{Adler:2002hy,Adler:2004hp,Anisimovsky:2004hr}.

While coupling the axion to electrons is desirable, an analogous coupling to muons with strength $\mathcal{O}(m_\mu/f_a)$ would induce contributions to $(g-2)_\mu$ that would violate present bounds unless $f_a\gsim v_{\text{EW}}$. Therefore, in this work we shall restrict ourselves to variant axion models that couple exclusively the first generation fermions, namely, $u$, $d$, and $e$.

\subsection{Pion-Phobia}
\label{Pion-Phobia}

The mixing of the axion with the neutral pion poses a major challenge to the viability MeV axion models. Severe constraints on this mixing have been placed three decades ago, and the idea of avoiding them via {\it pion-phobia} is just as old \cite{Krauss:1987ud}.

Axion-pion mixing induces the rare decay process $\pi^+\rightarrow e^+\nu_e a\rightarrow e^+\nu_e e^+e^-$.
The width is given by \cite{Krauss:1986bq}:
\be\label{piaenu}
\Gamma(\pi^+\rightarrow e^+\nu_e a)=\frac{\cos^2\theta_c}{384\pi^3}\,G_F^2\,m_\pi^5\;\theta_{a\pi}^{\,2}\,,
\ee
where $\theta_{a\pi}$ is the axion-pion mixing angle, and $\theta_c$ is the Cabibbo angle. The SINDRUM collaboration \cite{Eichler:1986nj} searched for this specific decay, and put bounds on $\Br(\pi^+\rightarrow e^+\nu_e (a\rightarrow e^+e^-))$ ranging from $(0.5-1)\times10^{-10}$ in the mass range $m_a\sim(1-20)\,\MeV$. Using (\ref{piaenu}), this translates into a severe upper limit on $\theta_{a\pi}$:
\be\label{api-limit}
|\,\theta_{a\pi}|\lsim(0.5-0.7)\times10^{-4}\,.
\ee
As we shall see in Sec.~\ref{dolor}, typical MeV axion variants are in conflict with (\ref{api-limit}), since they predict $\theta_{a\pi}\sim\mathcal{O}(f_\pi/f_a)\sim(0.5-10)\times10^{-2}$. As a consequence, the experimental upper bound on $\theta_{a\pi}$ pushes viable models into special regions of parameter space where the axion is {\it pion-phobic}, and therefore has suppressed couplings to isovector currents.

\subsection{Non-Robust/Uncertain Constraints}

Finally, some constraints are plagued by large hadronic uncertainties, and until those are under better control no robust exclusion claim can be made. That is the case of $a-\eta$ and $a-\eta^\prime$ mixings, which provide the dominant contribution to $K^+\rightarrow\pi^+a$.

Unlike $K^+\rightarrow\pi^+(a\rightarrow\text{invisible})$, the rare decay $K^+\rightarrow\pi^+(a\rightarrow e^+e^-)$ is much less constrained, $\Br\left(K^+\rightarrow\pi^+(a\rightarrow e^+e^-)\right)\lsim \mathcal{O}(10^{-6})$. In light of the allowed range (\ref{api-limit}) for $\theta_{a\pi}$, the contribution to this decay from $\pi^0-a$ mixing easily satisfies the experimental bound. However, it has been claimed in the literature \cite{Antoniadis:1981zw,Bardeen:1986yb} that the contribution to this amplitude from $a-\eta$ mixing is octet ($\Delta I=1/2$) enhanced, and translates into a bound $\theta_{a\eta}\lsim\mathcal{O}(10^{-4})$. Based on na\"{\i}ve estimates of $\theta_{a\eta}$ from leading order in chiral pertubation theory, it was then concluded that axions in the MeV mass range were hopelessly ruled out by $K^+$ decay bounds. In Sec.~\ref{kaons}, we will revisit these statements and constraints, and argue that due to the uncertainties involved in these estimates, previous claims of exclusion were overstated, and bounds from $K^+$ decays alone cannot definitively rule out the class of axion variants we consider.

Table~\ref{SummaryTable} summarizes the main experimental constraints relevant for generic MeV axions, and also lists the reasons as to why the variant model we shall introduce avoids all present bounds. This is not a standalone table - we urge the reader to follow our arguments in the main body of the paper to become fully aware of all the assumptions and caveats implicit in the information contained in Table~\ref{SummaryTable}.


\begin{table}
\begin{center}
\resizebox{1 \textwidth}{!}{ 
\renewcommand{\arraystretch}{1.4}
    \begin{tabular}{|c|c|c|c|c|}
	\hline
experimental limit & translates into & expectation for generic &expectation for MeV axion&\multirow{2}{*}{~reason~}\\[-0.3cm]
on observable & bound on model & MeV axions (LO in $\chi$PT) &  variant in (\ref{Laqq}) (NLO in $\chi$PT) & \\
	\botrule
\multirow{2}{*}{beam dumps  (see Fig.\ref{MeVaxionPlot})} & $\tau_a\lesssim 10^{-13}$ s  & $\tau_a< 10^{-12}$ s ~~if~~ $|\Qpq_e|\sim\mathcal{O}(1)$ & $|\Qpq_e|\sim\mathcal{O}(1)$ & \multirow{2}{*}{MB} \\
	  &  (see Fig.\ref{MeVaxionPlot}) & $\!\!\!\!\!\!\!\!\!\tau_a\gtrsim 10^{-11}$ s ~~if~~ $\Qpq_e=0$ & $\tau_a\lesssim 10^{-13}$ s~~for~~$m_a\gtrsim5-10$ MeV & \\	  
	\hline
	$~\Br(J/\Psi\rightarrow\gamma a)\lesssim 6\times10^{-2}$~ & $|\Qpq_c|\lesssim 0.25\,\Big(\frac{f_a}{\text{GeV}}\Big)$ & $|\Qpq_c|\sim\mathcal{O}(1)$ & $\Qpq_c=0$ & MB \\
	\hline
	$~\Br(\Upsilon\rightarrow\gamma a)\lesssim 3\times10^{-4}$~ & ~$|\Qpq_b|\lesssim 0.8\times10^{-2}\,\Big(\frac{f_a}{\text{GeV}}\Big)$~ & $|\Qpq_b|\sim\mathcal{O}(1)$ & $\Qpq_b=0$ & MB \\
	\hline
	$-\Delta(g-2)_\mu<0.5\times10^{-8}$ & $|\Qpq_\mu|\lesssim 4\times10^{-3}\,\Big(\frac{f_a}{\text{GeV}}\Big)$ & $|\Qpq_\mu|\sim\mathcal{O}(1)$ & $\Qpq_\mu=0$ & MB \\
	\hline
	\multirow{2}{*}{$-\Delta(g-2)_e<0.5\times10^{-11}$} & $|\Qpq_e|\lesssim \mathcal{O}$(few) & $|\Qpq_e|\sim\mathcal{O}(1)$ & $|\Qpq_e|\sim\mathcal{O}(1)$ & MB \\
	& $g_{a\gamma\gamma}\lesssim\mathcal{O}(0.1)\,\text{TeV}^{-1}$ & $g_{a\gamma\gamma}\sim\mathcal{O}(0.1-1)\,\text{TeV}^{-1}$& $g_{a\gamma\gamma}\sim\mathcal{O}(0.01-0.1)\,\text{TeV}^{-1}$&PP, NLO\\
	\hline
	$\Br(\pi^+\!\rightarrow e^+\nu(a\rightarrow e^+e^-))\lesssim 10^{-10}$ & $|\,\theta_{a\pi}|\lsim (0.5-0.7)\times10^{-4}$ & \multirow{2}{*}{$|\,\theta_{a\pi}|\sim\mathcal{O}(0.01-0.1)\Big(\frac{\text{GeV}}{f_a}\Big)$} & \multirow{2}{*}{$\theta_{a\pi}\sim(0.2\pm3)\times10^{-3}\Big(\frac{\text{GeV}}{f_a}\Big)$} & \multirow{2}{*}{PP, NLO} \\
	\cline{1-2}
	$\Delta\Br(\pi^0\rightarrow e^+e^-))\lesssim 2\times10^{-8}$ & $\Qpq_e\!\times\!\theta_{a\pi}\lsim 1.6\times10^{-4}\,\Big(\frac{f_a}{\text{GeV}}\Big)$ & & &\\
	\hline
\multirow{2}{*}{$\Br(K^+\!\rightarrow \pi^+(a\rightarrow e^+e^-))\lesssim 10^{-5}-10^{-6}$} & $|\,\theta_{a\eta_{ud}}|\lsim 10^{-4}$ if octet enhanced& \multirow{2}{*}{~$|\,\theta_{a\eta_{ud}}|\sim\mathcal{O}(10^{-3}-10^{-2})\Big(\frac{\text{GeV}}{f_a}\Big)$~} & \multirow{2}{*}{~$\theta_{a\eta_{ud}}\sim(-2\pm8)\times10^{-3}\Big(\frac{\text{GeV}}{f_a}\Big)$~} & \multirow{2}{*}{NLO} \\
	 &\!\!\!\!\!\!\!\!\!\!\!\!\!\!\!\!\!\!\!\!\!\!\!\!\!\!\!\!\!\!\!\!\! $|\,\theta_{a\eta_{ud}}|\lsim 10^{-2}$ if not & &  &  \\
	\hline
	$\Br(K^+\!\rightarrow \pi^+(a\rightarrow\gamma\gamma))\lesssim 10^{-9}$ & \multirow{2}{*}{$|\,\theta_{a\eta_{s}}|\lsim \mathcal{O}(10^{-1})$} & \multirow{2}{*}{$|\,\theta_{a\eta_{s}}|\sim \mathcal{O}(10^{-2})\Big(\frac{\text{GeV}}{f_a}\Big)$} & \multirow{2}{*}{$\theta_{a\eta_{s}}\sim (1\pm2)\times10^{-2}\Big(\frac{\text{GeV}}{f_a}\Big)$} & \multirow{2}{*}{NLO} \\
	$\Br(\Phi\!\rightarrow \gamma(a\rightarrow e^+e^-))\lesssim 5\times10^{-5}$ & & & & \\
	\hline
	\multirow{2}{*}{$\Br(K^+\!\rightarrow \pi^+(a\rightarrow \text{inv}))\lesssim 0.5\times10^{-10}$} & $\Br(a\rightarrow \text{inv})\lesssim \mathcal{O}(10^{-4})$ & \multirow{2}{*}{$\Br(a\rightarrow \nu\bar\nu)\sim({\Qpq_\nu} m_\nu/\Qpq_e m_e)^2$} & $\Br(a\rightarrow \chi\bar\chi)\sim({\Qpq_\chi} m_\chi/\Qpq_e m_e)^2$ & \multirow{2}{*}{MB} \\
	 & (assuming $\tau_a< 10^{-12}$ s) &  & $\chi\,=\,\nu_{(e,\mu,\tau,s)}$, sub-MeV DM, ...  &  \\
	\hline
    \end{tabular}
    \renewcommand{\arraystretch}{1}
    }
\end{center}
\caption{\scriptsize Summary of the most relevant existing bounds on MeV axions, as well as the conditions for experimental viability discussed in this paper. The 3rd column indicates the typical range in generic axion models for the quantity being constrained, estimated at leading order (LO) in chiral perturbation theory ($\chi$PT). The 4th column contrasts it with the corresponding expectation for a short-lived, {\it pion-phobic} MeV axion which, as we argue in the text, remains compatible with experimental bounds. The 4th column takes into account large uncertainties from next-to-leading order (NLO) in $\chi$PT. The fifth column indicates the reason for the difference in expected properties of generic models (3rd column) and of the viable variant (4th column), where: {\bf MB} indicates that the quantity in the 4th column is achievable via model building; {\bf NLO} indicates that the error in the quantity in the 4th column comes from large corrections at NLO in $\chi$PT; and {\bf PP} indicates that the quantity in the 4th column is suppressed due to accidental cancelations at LO in the axion-pion mixing angle $\theta_{a\pi}$, i.e., {\it pion-phobia}.}
\label{SummaryTable}
\end{table}

\section{\label{dolor} Axion-Meson Mixings in $\chi$PT @ LO}


We can trivially evade the constraints discussed in Sec.~\ref{avoidable} by restricting our investigation to {\it QCD axion variants coupled only to the first generation of SM fermions}. This assumption is implied for the remainder of this paper.

We are now in a position to introduce more concrete notation and review the extraction of axion-meson mixing angles to leading order in chiral perturbation theory ($\chi$PT).

We define the axion couplings slightly above the QCD scale, where all heavy fermions ($c$, $b$ and $t$) have been integrated out. We also choose a specific basis (motivated by UV completions to be discussed in Sec.\,\ref{GeVcompletion}) in which the operators $(a/f_a)\,G\tilde{G}$ and $\partial_\mu (a/f_a)J^{5\,\mu}_{\text{SM}}$ are not present, or have been rotated away. With this choice of basis, the variant models we shall consider are unambiguously defined by the following couplings:
\be
\label{Laqq}
\mathcal{L}_a^{\text{eff}}~ =~ m_u\,e^{i\,\Qpq_u\,a/f_a}\,uu^c\;+\; m_d\,e^{i\,\Qpq_d\,a/f_a}\,dd^c 
\;+\;m_e\,e^{i\,\Qpq_e\, a/f_a}\,ee^c \;+\; \text{h.c.}
 \ee 
 
Above, $\Qpq_f$ are Peccei-Quinn (PQ) charges that determine the transformation of fermions $f = u,d,e$ under a PQ rotation.
Below the QCD scale, (\ref{Laqq}) can be mapped onto the effective chiral Lagrangian by treating the axion as a complex phase of the quark masses. To leading order in $\chi$PT, the axion couplings to the $\U(3)_\chi$ meson nonet can be extracted from:
\be
\label{chiL0}
\mathcal{L}_{\chi}^{(0)} = \frac{f_\pi^2}{4} \, \Tr\left[ 2B M_q(a)\, U+ \text{h.c.} \right]\;-\; \frac{1}{2}M_0^2\eta_0^2\, ,
\ee
where $f_\pi=92$ MeV,
$B$ has mass dimensions and is set by the QCD quark condensate, $M_0\sim\mathcal{O}(\GeV)$ parameterizes the strong anomaly contribution to the mass of the $\SU(3)_\chi$ singlet $\eta_0$, $M_q(a)$ is the light quark mass matrix\footnote{From this point forward we will suppress the superscript in the PQ charges, $\Qpq_f\rightarrow Q_f$, to lighten the notation.},
\be\label{Mqa}
M_q(a)\equiv
\begin{pmatrix} 
\,m_u\,e^{i\,Q_u a/f_a} & & \\
  & m_d\,e^{i\,Q_d\,a/f_a}  &    \\
  &  & m_s~
\end{pmatrix},
\ee
and $U$ is the non-linear representation of the meson nonet under chiral $\U(3)_\chi$ symmetry
\be
U\equiv \text{Exp}\;i\frac{\sqrt{2}}{f_\pi}\;
\begin{pmatrix} 
\vspace{0.17cm}
\frac{\pi^0}{\sqrt{2}}+\frac{\eta_8}{\sqrt{6}}+\frac{\eta_0}{\sqrt{3}} & \pi^+ & K^+ \\
\vspace{0.17cm}
\pi^-  & -\frac{\pi^0}{\sqrt{2}}+\frac{\eta_8}{\sqrt{6}}+\frac{\eta_0}{\sqrt{3}}  &  K^0  \\
  K^- & \overline{K}^0 & -\frac{\eta_8}{\sqrt{3/2}}+\frac{\eta_0}{\sqrt{3}}~
\end{pmatrix}.
\ee

Expanding the leading order chiral Lagrangian (\ref{chiL0}) to quadratic terms in the fields and diagonalizing the mass matrix, we obtain:
\beq
m_a^2 &=& (Q_u+Q_d)^2\,\frac{B\; m_u m_d m_s}{\big(m_u m_d+m_u m_s+m_d m_s+\frac{6\, B\, m_u m_d m_s}{M_0^2}\big)}\,\frac{f_\pi^2}{f_a^2} \label{theta-a-pi}\nonumber\\
 &=& \frac{(Q_u+Q_d)^2}{(1+\epsilon_{\eta\eta^\prime})}\,\frac{m_um_d}{(m_u+m_d)^2}\,\frac{m_\pi^2\, f_\pi^2}{f_a^2}\; ,\label{ma}
\eeq
with
\beq
\epsilon_{\eta\eta^\prime}&\equiv&\frac{m_um_d}{m_s(m_u+m_d)}\;\bigg(1+6\,\frac{B\, m_s}{M_0^2}\bigg)\nonumber\\
&\approx&\frac{m_um_d}{(m_u+m_d)^2}\;\frac{m_\pi^2}{m_K^2}\;\bigg(1+6\,\frac{m_K^2}{m_{\eta^\prime}^2}\bigg)~\simeq~0.04\;,
\eeq
and
\beq
\theta_{a\pi}^{(0)} &=& -\frac{1}{(1+\epsilon_{\eta\eta^\prime})}\;\bigg(\frac{(Q_u m_u-Q_d m_d)}{(m_u+m_d)}+\epsilon_{\eta\eta^\prime}\,\frac{(Q_u-Q_d)}{2}\bigg)\,\frac{f_\pi}{f_a}\;, \label{theta-a-pi}\\
\theta_{a\eta_{_8}}^{(0)} &=& -\frac{\sqrt{3}}{~2}\;\frac{\epsilon_{\eta\eta^\prime}}{(1+\epsilon_{\eta\eta^\prime})}\;\frac{\Big(1+\frac{2\,B\, m_s}{M_0^2}\Big)}{\Big(1+\frac{6\,B\, m_s}{M_0^2}\Big)}\;(Q_u+Q_d)\; \frac{f_\pi}{f_a}\;, \label{theta-a-8}\\
\theta_{a\eta_{_0}}^{(0)} &=&-\sqrt{6}\;(Q_u+Q_d)^{-1}\; \frac{f_a}{f_\pi}\;\frac{m_a^2}{M_0^2}\;,\label{theta-a-0}
\eeq
where the superscript $^{(0)}$ on the mixing angles indicates that these expressions stem from leading order in $\chi$PT. Note, in particular, that when any of the quark masses is taken to zero, both the axion mass, $m_a$, and its mixing with the SU(3)$_\chi$ singlet, $\theta_{a\eta_{_0}}$, vanish. This is consistent with the fact that, in the limit of a massless quark, the axion becomes a {\it bona fide} Goldstone boson associated with a non-anomalous conserved current.

Previous studies in the literature have assumed that these leading order estimations of the axion-meson mixing angles were good enough approximations. Based on those, they went on to extract rates for processes involving the axion, and infer bounds that widely excluded all axion variants with $m_a\gtrsim\mathcal{O}$(MeV). The core argument of this paper is that these assumptions, while correct for a wide class of axion variants, fail under special circumstances and invalidate broad exclusion claims. In these special circumstances, the mixing angles in (\ref{theta-a-pi}-\ref{theta-a-0}) receive $\mathcal{O}$(1) corrections from the next order in the chiral expansion, introducing large uncertainties to the axion couplings and dramatically affecting the inference of bounds that depend on these mixing angles.


\section{\label{dolor} Constraints from Pion Decays}

As discussed in Sec.~\ref{Pion-Phobia}, bounds from $\pi^+\rightarrow e^+\nu_e\, a$ severely restrict viable MeV axions to the extreme {\it pion-phobic} region. In this section we will discuss under what conditions {\it pion-phobia} can be achieved, as well as implications for another relevant pion decay, namely, $\pi^0\rightarrow e^+e^-$.

The leading order expression for the axion mixing with the neutral pion in (\ref{theta-a-pi}) indicates that, for generic $\mathcal{O}(1)$ values of $Q_{u,d\,}$, $\theta_{a\pi}\sim\mathcal{O}(f_\pi/f_a)$. For this generic situation, the axion-pion mixing is about one to two orders of magnitude larger than the other axion-meson mixings, and therefore dominates the axion phenomenology involving hadronic couplings. 

However, notice that $|\theta_{a\pi}^{(0)}|$ in (\ref{theta-a-pi}) is not bounded from below. 
In particular, for
\be\label{thetadeltaz}
\frac{Q_u}{Q_d}=2\quad\Rightarrow\quad \theta_{a\pi}^{(0)}\approx\frac{4\,Q_d}{3}\,\frac{f_\pi}{f_a}\left(\frac{1}{2}-\frac{m_u}{m_d}\right)\approx0.
\ee
Indeed, plugging in the PDG's 2017 weighted avarage for the light quark mass ratio \cite{Patrignani:2016xqp}, $m_u/m_d=0.483\pm0.027$, and setting $Q_d=1$ without loss of generality, (\ref{theta-a-pi}) gives:
\be\label{thetarange}
\theta_{a\pi}^{(0)}~\approx~\frac{(0.2\pm3)\times10^{-3}}{f_a/\GeV},
\ee
which contains the {\it pion-phobic} range (\ref{api-limit}) still allowed by $\pi^+\rightarrow e^+\nu_e\,a$.

Therefore, extreme {\it pion-phobia} might be achievable in variant models with $Q_d/Q_u=1/2\approx m_u/m_d$. Unfortunately, under these circumstances, the leading order expression (\ref{theta-a-pi}) is not a reliable estimate of $\theta_{a\pi}$.
The reason is that the second order expansion in $\chi$PT can give comparable contributions to $\theta_{a\pi}$, which should be included for a reliable estimate of this quantity. However, since the coefficients of many $\mathcal{O}(p^4)$ $\chi$PT operators that contribute to $\theta_{a\pi}$ are poorly known,
 a precise determination of $\theta_{a\pi}$ is not possible. Moreover, as we will discuss in Sec.~\ref{GeVcompletion}, additional GeV states from the UV completion of such models also give model dependent contributions to $\theta_{a\pi}$.

In summary, we have identified a particular class of axion variants, defined in (\ref{Laqq}) with $Q_u/Q_d=2$, which is compatible with the condition of extreme {\it pion-phobia}, and hence remains a viable possibility. Due to present errors in the determination of $m_u/m_d$, uncertainties in the second order expansion in $\chi$PT, and model dependence of associated UV completions, the estimated range for the axion-pion mixing (\ref{thetarange}) is much broader than the experimentally allowed range in (\ref{api-limit}), but it is nonetheless consistent with those bounds given errors.

\subsection{The KTeV anomaly}
\label{KTeVanomaly}

Were such {\it pion-phobic} axion to exist, $\theta_{a\pi}$ would ideally be determined from experiment. Another very sensitive probe of this mixing angle is the $\pi^0$ decay width to $e^+e^-$, which is highly suppressed in the Standard Model. In fact, there appears to be a discrepancy in the observed $\pi^0$ decay width to $e^+e^-$ at the $\sim 3\sigma$ level. The most precise measurement of this branching ratio, from the KTeV-E778 collaboration \cite{Abouzaid:2006kk}, is:
\be\label{ktev}
\Br(\pi^0\rightarrow e^+e^-)\big|_{\text{KTeV}} = (7.48\pm0.29\pm0.25)\times10^{-8},
\ee
where the first and second errors are statistical and systematic, respectively. Theoretical estimates of this branching ratio \cite{Dorokhov:2007bd,Dorokhov:2008qn,Dorokhov:2009xs} in the SM predict:
\be\label{SMBr}
\Br(\pi^0\rightarrow e^+e^-)\big|_{\text{SM}} = (6.23\pm0.09)\times10^{-8},
\ee
corresponding to a $3.2\sigma$ discrepancy between theory and experiment\footnote{The KTeV-E778 collaboration extrapolated its measurement in the exclusive region $m_{ee}^2/m_\pi^2>0.95$ to obtain the inclusive result with final state radiation removed \cite{Abouzaid:2006kk}. Refs. \cite{Vasko:2011pi,Husek:2014tna} revisited the QED radiative corrections used in this extrapolation, and claimed that the discrepancy between theory and experiment could be reduced to the $2\sigma$ level.}.

In the SM, this decay is loop induced via the $\pi^0\gamma\gamma$ coupling. The imaginary (absorptive) part of this amplitude is well understood and gives an irreducible contribution, the so called ``unitarity bound'', to this branching ratio \cite{Landsberg:1986fd}:
\be\label{ImBr}
\Br(\pi^0\rightarrow e^+e^-)\big|_{\text{unit.}}\,=~\Br(\pi^0\rightarrow \gamma\gamma)\,\frac{\alpha^2}{2\beta}\,\log^2\left(\frac{1+\beta}{1-\beta}\right)\,\frac{m_e^2}{m_\pi^2}~\approx~ 4.7\times10^{-8}\,,
\ee
where $\beta\equiv\sqrt{1-4m_e^2/m_\pi^2}$. By noting that the width for $\pi^0\rightarrow\gamma\gamma$ is given by:
\be
\Gamma(\pi^0\rightarrow\gamma\gamma)=\bigg(\frac{\alpha}{4\pi f_\pi}\bigg)^2\; \frac{m_\pi^3}{4\pi},
\ee
we can model the imaginary part of the SM amplitude $\text{Im}\mathcal{M}(\pi^0\rightarrow e^+e^-)_\text{SM}$ via an effective $\pi^0e^+e^-$ vertex given by \cite{Bergstrom:1982wk}:
\be
\mathcal{L}_{\pi e e}\supset i\, y^{\text{SM}}_{\pi ee}\,\pi^0\,\overline{e}\gamma^5e
\quad ,\quad
y^{\text{SM}}_{\pi ee} \equiv \frac{1}{2\beta}\,\log\left(\frac{1+\beta}{1-\beta}\right)\,\frac{\alpha^2}{2\pi}\,\frac{m_e}{f_\pi}.
\ee
Parameterizing BSM corrections to this coupling as $y^{\text{BSM}}_{\pi ee}$, we can finally write
\be
\mathcal{L}_{\pi e e} = i\,\left( y^{\text{SM}}_{\pi ee}+y^{\text{BSM}}_{\pi ee}\right)\,\pi^0\,\overline{e}\gamma^5e\,,
\ee
and predict the contribution to the width from the imaginary part of the amplitude:
\be\label{widthpiee}
\Gamma(\pi^0\rightarrow e^+e^-)\big|_{\text{Im}\mathcal{M}} ~=~ \beta\,\frac{m_\pi}{8\pi}\, \left| y^{\text{SM}}_{\pi ee}+y^{\text{BSM}}_{\pi ee}\right|^2.
\ee

As a conservative bound on the BSM contribution to $\Gamma(\pi^0\rightarrow e^+e^-)$, we can demand that (\ref{widthpiee}) does not exceed the observed value (\ref{ktev}) by more than two standard deviations, leading to:
\be\label{KTeVboundY}
-6.0\times 10^{-7}~ ~\lsim~~ y^{\text{BSM}}_{\pi ee}~~ \lsim~~ 0.83\times10^{-7}\,.
\ee
Note that when $y^{\text{BSM}}_{\pi ee}<0$, the BSM amplitude destructively interferes with the SM one, and in the range $y^{\text{BSM}}_{\pi ee}\in [-5.2,0]\times10^{-7}$ the contribution (\ref{widthpiee}) to the width from the imaginary part of the amplitude is smaller than the SM unitarity bound (\ref{ImBr}). In that case, the real (dispersive) part of the amplitude would have to be unexpectedely large to account for the observed width.

If instead we attempt to explain the excess\footnote{See also \cite{Kahn:2007ru} for an alternative BSM explanation of the KTeV anomaly.}, i.e., fit the BSM contribution to account for the discrepancy between (\ref{ktev}) and (\ref{SMBr}), we obtain the following $2\sigma$ range:
\be
y^{\text{BSM}}_{\pi ee} \Big|_\text{KTeV}= 
\begin{cases}
~~(0.3\pm0.2)\times10^{-7}\quad\text{or}\\
(-5.5\pm0.2)\times10^{-7}.
\end{cases}
\ee

So far we have kept this discussion general, since for some UV completions of the axion variants we are considering, not only the axion, but also other GeV states, can contribute to $y^{\text{BSM}}_{\pi ee}$. Nevertheless, we can consider the limit of these models in which the only important contribution to $y^{\text{BSM}}_{\pi ee}$ comes from the QCD axion. In this case,
\be
y^{\text{BSM}}_{\pi ee} =\theta_{a\pi}\times\frac{ Q_e\,m_e}{f_a}\,.
\ee
The $2\sigma$ bound (\ref{KTeVboundY}) in this instance then becomes:
\be\label{KTeVboundTheta}
\theta_{a\pi}~\lsim~\frac{1.6\times10^{-4}}{Q_e\,(\GeV/f_a)},
\ee
and the range needed to fit the KTeV anomaly within 2 standard deviations is:
\be\label{a-pi-KTeV}
\theta_{a\pi}\big|_\text{KTeV}\;\approx\;\frac{(0.6\pm0.4)\times10^{-4}}{Q_e\,(\GeV/f_a)}.
\ee
Remarkably, the $\theta_{a\pi}$ fit to the KTeV anomaly, (\ref{a-pi-KTeV}), is compatible with the {\it pion-phobic} range imposed by $\pi^+\rightarrow e^+\nu_e\,a$.

\section{\label{kaons} Constraints from Charged Kaon Decays}

It was widely claimed in the literature \cite{Antoniadis:1981zw,Bardeen:1986yb} that bounds from $K^+\rightarrow\pi^+ a$ ruled out all QCD axion variants in the MeV mass range. In this section we critically re-examine the arguments which have led to this claim. We find that there are large uncertainties involved in obtaining $K^+\rightarrow\pi^+ a$, from the interpretation of experimental analyses, to assumptions in estimating the $K^+\rightarrow \pi^+(\eta^*\rightarrow a)$ amplitude, and finally to the derivation of the $\eta-a$ mixing angle. We conclude that these uncertainties are significant enough to preclude a definitive exclusion of MeV axions from existing bounds on rare $K^+$ decays.

\subsection{Experimental bounds on $K^+\rightarrow\pi^+ (a\rightarrow e^+e^-)$}
\label{Kexp}

Measurements of rare $K^+$ decays such as $K^+\rightarrow\pi^+ \nu\overline{\nu}$, $K^+\rightarrow\pi^+ \gamma\gamma$, and $K^+\rightarrow\pi^+ e^+e^-$ have been improving over time for the past five decades or so. However, for the last 30 years, the low $m_{e^+e^-}$ region of $K^+\rightarrow\pi^+ e^+e^-$ has been neglected from experimental scrutiny for BSM physics.
This is due in part to the very large backgrounds from $\pi^0$ Dalitz decays, namely, $K^+\rightarrow\pi^+(\pi^0\rightarrow e^+e^-\gamma)$, which make the study of the low $m_{e^+e^-}$ region very difficult. To the best of our knowledge, the last time dedicated searches were performed in this region was in the 1980's, by two different experiments, one at KEK \cite{Yamazaki:1984vg} and the other at BNL \cite{Baker:1987gp}. 
We shall now discuss these searches in some detail.

The KEK experiment E89 \cite{Yamazaki:1984vg} was published in 1984. It consisted of a high resolution spectrograph that measured the momentum of charged pions from the decay of charged kaons at rest. They searched for a peak in the $\pi^+$ momentum distribution, which would be evidence of a two-body decay of the charged kaon, $K^+\rightarrow\pi^+X^0$, to a new pseudoscalar $X^0$. Since they did not impose any vetoes nor requirements on the remaining decay products of $K^+$, this search was sensitive to $X^0\rightarrow\text{anything}$. Unfortunately, we believe there are a couple of issues with this study.

First, it is not clear what was the lowest value of $m_{X^0}$ to which this search was sensitive. The abstract states that ``{\it bounds are presented for the mass range of $X^0$ from 10 to 300 MeV/$c^2$.}'' The concluding paragraph, on the other hand, quotes $5-300~\MeV$ as the range of exclusion for $m_{X^0}$. The acceptance curve for the pion momentum, shown in Fig.~1 of \cite{Yamazaki:1984vg}, 
has a lower range of $P_{\pi^+}\sim 224~\MeV$, corresponding to a lower range of $50~\MeV$ for $m_{X^0}$. Finally, the exclusion curve presented in Fig.~2, of $\Br(K^+\rightarrow\pi^+X^0)$ versus $m_{X^0}$, has a lower range of $m_{X^0}=10~\MeV$.

Moreover, there are issues with Fig.~2. It misrepresents limits from previous analysis, namely, Asano et al.\cite{Asano:1982fq} and Abrams et al.\cite{Abrams:1977ex}. The limits from Asano et al.~on $\Br(K^+\rightarrow\pi^+ (X^0\rightarrow \gamma\gamma)\,)$ are depicted in Fig.~2 as a factor of $\sim 5$ weaker than in \cite{Asano:1982fq}, and shown only in the range $m_{X^0}\in(50-90~\MeV)$, when in fact Asano et al.'s limits apply from $(0-100~\MeV)$ as long as $\tau_{X^0}<10^{-9}$~s. It is unclear which limit from Abrams et al. \cite{Abrams:1977ex} is being depicted in Fig.~2, but assuming it is the limit on $\Br(K^+\rightarrow\pi^+ \gamma\gamma)$ in the region $K_{\pi^+}<70~\MeV$, corresponding to $P_{\pi^+}> 156~\MeV$ ($M_{\gamma\gamma}>237~\MeV$), then Fig.~2 of \cite{Yamazaki:1984vg} misrepresents it by a factor of $\sim5$ as well.

An earlier conference note \cite{Yamazaki:1984qx} presenting preliminary results of the KEK E89 analysis might shed light on the origin of these discrepancies\footnote{We also attempted to contact some of the authors of \cite{Yamazaki:1984vg}, but since three decades have elapsed since the time of publication of this analysis, we understandably did not get a response to our queries.}. In Fig.~6 of conference note \cite{Yamazaki:1984qx}, limits were presented in terms of the relative branching ratio $\Br(K^+\rightarrow\pi^+X^0)/\Br(K^+\rightarrow\pi^+\pi^0) \approx 5\times\Br(K^+\rightarrow\pi^+X^0)$, and represented more accurately the results from Asano et al.\cite{Asano:1982fq} and Abrams et al.\cite{Abrams:1977ex}. This suggests that the factor of $\sim 5$ discrepancy in the final publication \cite{Yamazaki:1984vg} was a typo in recasting the plot from relative branching ratio to absolute branching ratio. Adding to the ambiguity on the lowest ${X^0}$ mass to which KEK E89 was sensitive, the limits presented on Fig.~6 of the conference note \cite{Yamazaki:1984qx} did not extend below $m_{X^0}\sim50~\MeV$.

The second issue with the KEK E89 analysis was their statistical inference of exclusion limits in the low statistics region $m_{X^0}\lsim 80~\MeV$. The formula used for obtaining limits is appropriate in the case of a ``bump-hunt'' on top of a very large background, where the errors are gaussian and scale as $\sqrt{N_\text{background}}$. It fails, however, in the Poisson limit where the number of expected events from signal plus background is $\mathcal{O}(1)$, where it can overestimate the exclusion power by a factor of several.

All things considered, we conservatively choose to ignore the limits from KEK E89 analysis \cite{Yamazaki:1984vg} in the region $m_{X^0}\lsim 50~\MeV$.

That leaves us with the other dedicated search for the rare decay $K^+\rightarrow\pi^+ (a\rightarrow e^+e^-)$ in the low $m_{e^+e^-}$ region, published in 1987 by Baker et al. \cite{Baker:1987gp}. Their apparatus, located at BNL's AGS, was optimized for the decay of $K^+$ to three charged tracks, with efficient discrimination between electrons, pions and muons. They collected a large sample of approximately $2.8\times10^4$ $K^+\rightarrow\pi^+ e^+e^-$ events, mostly Dalitz decays, $K^+\rightarrow\pi^+(\pi^0\rightarrow e^+e^-\gamma)$, and applied optimized cuts to remove this background while remaining sensitive to the $K^+\rightarrow\pi^+ a$ signal.  While they claim that their results exclude branching ratios $\Br(K^+\rightarrow\pi^+ a)\gsim0.8\times10^{-7}$ for any mass in the range $m_a\in(1.8-100~\MeV)$ at 90\% confidence level, we believe their exclusion is too aggressive for two reasons.

Firstly, in their final signal region, they fit the observed  $m_{e^+e^-}$ distribution in the range $1.8-100~\MeV$ as a constant background not properly modeled by their Monte Carlo (MC), and then subtract it before extracting limits. We believe that a more conservative approach when deriving limits would be to treat any residual events in the signal region as potential signal. We can estimate what those more conservative limits would be based on the data shown in Fig.~2(d) of \cite{Baker:1987gp}, together with the assumption that the signal efficiency as a function of $m_a$ does not vary substantially from the one quoted at $m_a=1.8~\MeV$, and trusting their Monte Carlo signal yield of 5 expected events for $\Br(K^+\rightarrow\pi^+a)=10^{-6}$. Under these assumptions, we infer the limit $\Br(K^+\rightarrow\pi^+ a)\lsim2.4\times10^{-6}$ in the range $m_a\sim(10-25)~\MeV$.

Our second concern regards this analysis' sole reliance on Monte Carlo to estimate all the acceptances of both background and signal. As Fig.~2(d) of \cite{Baker:1987gp} shows, their Monte Carlo grossly misestimates the Dalitz background contamination in the signal region by about one order of magnitude. They do not address this mismodeling and simply re-scale their expected background to match the observed rate. More importantly, they do not address whether their Monte Carlo estimation of the {\it signal acceptance} could possibily be mismodeled as well. Were their signal acceptance off by a similar order of magnitude, their limits could be weakened to as much as $\Br(K^+\rightarrow\pi^+ a)\lsim10^{-5}$ in the low $m_{e^+e^-}$ region.

We therefore choose to remain agnostic about the precise experimental limit on MeV axions from kaon decays, and for the remainder of this paper we discuss implications from $K^+$ decays to axion-meson mixing parameters assuming the following range:
\be\label{Kexpbound}
\Br(K^+\rightarrow\pi^+a\rightarrow\pi^+ e^+e^-)\lsim 10^{-6}-10^{-5}.
\ee

We stress that new experimental studies of the low $m_{ee}$ kinematic region in $K^+\rightarrow\pi^+ e^+e^-$ are relevant and warranted, considering the sensitivity of this final state to rare new physics processes, and the fact that measurements in this region could be greatly improved by modern experiments.

\subsection{Estimation of the amplitude for $K^+\rightarrow\pi^+ a$}
\label{DeltaI=1/2}

Next, we discuss how to relate the amplitude for $K^+\rightarrow\pi^+ a$ to the SM amplitudes for $K^+\rightarrow\pi^+ \pi^0$, $K^+\rightarrow\pi^+\eta_{{8}}$, and $K^+\rightarrow\pi^+ \eta_{{0}}$, assuming that the axion-meson mixing angles $\theta_{a\pi}$, $\theta_{a\eta_{_{8}}}$, and $\theta_{a\eta_{_{0}}}$ are known.

Since the $\pi^0$ is on-shell in the SM process $K^+\rightarrow\pi^+ \pi^0$, it is trivial to relate the amplitude for this decay to the amplitude for $K^+\rightarrow\pi^+ a$ via axion-pion mixing:
\be\label{Kfromapi}
\mathcal{M}(K^+\rightarrow\pi^+ a)\big|_{a-\pi\, \text{mixing}}=~\theta_{a\pi}~\mathcal{M}(K^+\rightarrow\pi^+ \pi^0).
\ee
If (\ref{Kfromapi}) were the dominant contribution to $\mathcal{M}(K^+\rightarrow\pi^+ a)$, we could plug in the upper bound (\ref{api-limit}) on $\theta_{a\pi}$ to obtain:
\be
\Br(K^+\rightarrow\pi^+ a)\big|_{a-\pi\, \text{mixing}}~\lsim~(0.5\times10^{-8})\times\Br(K^+\rightarrow\pi^+ \pi^0)~\approx~10^{-9},
\ee
which is at least three orders of magnitude below the existing bounds discussed in Sec.~\ref{Kexp}.

Antoniadis et al.~\cite{Antoniadis:1981zw} and Bardeen et al.~\cite{Bardeen:1986yb} pointed out that the axion mixing with $\eta$ and $\eta^\prime$ also contribute to $\mathcal{M}(K^+\rightarrow\pi^+ a)$:
\be\label{Kfromaeta}
\mathcal{M}(K^+\rightarrow\pi^+ a)\big|_{a-\eta_{_{8,0}}\, \text{mixing}}~=~\theta_{a\eta_{_{8}}}\;\mathcal{M}(K^+\rightarrow\pi^+ \eta_8) ~+~ \theta_{a\eta_{_{0}}}\;\mathcal{M}(K^+\rightarrow\pi^+ \eta_0),
\ee
and that, due to {\it $\Delta I=1/2$ enhancement} (a.k.a. {\it octet enhancement}) of these amplitudes, these contributions dominate the branching ratio for $K^+\rightarrow\pi^+ a$ and lead to tension with experimental constraints.

Let us briefly review the arguments of \cite{Bardeen:1986yb}. In order to obtain the amplitudes $\mathcal{M}(K^+~\rightarrow~\pi^+ \eta_{8,0})$, \cite{Bardeen:1986yb} considers the leading order $\Delta S=1$ chiral Lagrangian describing  non-leptonic kaon decays:
\be\label{Loctet27plet}
\mathcal{L}_\chi^{\Delta S=1}\Big|_{\mathcal{O}(p^2)}\!\!=~g_8\,f_\pi^2\,\Tr\left(\lambda_{ds}\, \partial_\mu U\, \partial^\mu U^\dagger\right) ~+~ g_{27}\,f_\pi^2\,C_{ab}\,\Tr\left(\lambda_a\, \partial_\mu U\,U^\dagger\,\lambda_b\, \partial^\mu U\, U^\dagger\right)\,+\,\text{h.c.},
\ee
where $\lambda_a$ are the Gell-Mann matrices, $\lambda_{ds}\equiv(\lambda_6+i\lambda_7)/2$, and the Clebsch-Gordan coefficients $C_{ab}$ are given for instance in \cite{Donoghue:1992dd}. The first operator transforms as an octet under chiral $\SU(3)_\chi$, while the second transforms as a 27-plet. 
In particular, while $\mathcal{M}(K^0\rightarrow\pi\pi)$ receives contributions from both operators, only the 27-plet operator contributes to $\mathcal{M}(K^+\rightarrow\pi^+\pi^0)$:
\beq
\mathcal{M}(K^0\rightarrow\pi^+\pi^-)&=&\frac{\sqrt{2}}{f_\pi}(m_K^2-m_\pi^2)\left(g_8\,e^{i\delta_0}+g_{27}\,e^{i\delta_2} \right)\,,\label{K0toPi+Pi-}\\
\mathcal{M}(K^0\rightarrow\pi^0\pi^0)&=&\frac{\sqrt{2}}{f_\pi}(m_K^2-m_\pi^2)\left(g_8\,e^{i\delta_0}-2g_{27}\,e^{i\delta_2} \right)\,,\label{K0toPi0Pi0}\\
\mathcal{M}(K^+\rightarrow\pi^+\pi^0)&=&\frac{3}{f_\pi}(m_K^2-m_\pi^2)\,g_{27}\,e^{i\delta_2}\,.\label{K+toPi+Pi0}
\eeq
Above, $\delta_0,\,\delta_2$ are strong interaction S-wave $\pi\pi$ phase shifts.
Within the effective framework of $\chi$PT, the coefficients $g_8$ and $g_{27}$ cannot be obtained from first principles. However, under the standard assumption that (\ref{K0toPi+Pi-}, \ref{K0toPi0Pi0}, \ref{K+toPi+Pi0}) provide the leading contributions non-leptonic kaon decays, $g_8$ and $g_{27}$ can be fit to match the observed $K\rightarrow\pi\pi$ amplitudes \cite{Donoghue:1992dd} (see also \cite{Cirigliano:2011ny}):
\be\label{g8g27}
g_8\,\simeq \, 7.8\times 10^{-8}\,,\qquad g_{27}\,\simeq\, 0.25\times 10^{-8}\, .
\ee
Whereas na\"{\i}vely one would expect $g_8\sim g_{27}$, (\ref{g8g27}) shows a large enhancement of the octet coefficient relative to the 27-plet's,  ~$g_8/g_{27}\simeq 31.2$, ~tied to the fact that the hadronic width of $K^0_{S}$ is much broader than that of $K^\pm$, e.g.,
\be
\frac{\Gamma(K^0_S\rightarrow\pi\pi)}{\Gamma(K^+\rightarrow\pi^+\pi^0)} \approx 668\,.
\ee
This is known as the {\it octet enhancement} in non-leptonic kaon decays.

Ref.~\cite{Bardeen:1986yb} noted that the octet operator in (\ref{Loctet27plet}) also contributes to the (off-shell) amplitudes:
\beq
\mathcal{M}(K^+\rightarrow\pi^+\eta_8)&=&\frac{1}{\sqrt{3}\,f_\pi}(2\,m_K^2+m_\pi^2-3\,p^2_{\eta_8})\,g_8\,,\label{K+toPi+eta8}\\
\mathcal{M}(K^+\rightarrow\pi^+\eta_0)&=&\frac{2\sqrt{2}}{\sqrt{3}\,f_\pi}(m_K^2-m_\pi^2)\,g_8\,.\label{K+toPi+eta0}
\eeq

Defining the quark flavor basis:
\be
\eta_{_{ud}}~\equiv~\frac{\eta_8}{\sqrt{3}}+\frac{\eta_0}{\sqrt{3/2}}~~~~~,~~~~~\eta_s~\equiv-\frac{\eta_8}{\sqrt{3/2}}+\frac{\eta_0}{\sqrt{3}}\;,
\ee
and neglecting terms $\mathcal{O}(g_{27}/g_8)$ and $\mathcal{O}(m_\pi^2/m_K^2)$, we can then use (\ref{K+toPi+eta8},\ref{K+toPi+eta0}), (\ref{Kfromaeta}) and (\ref{K0toPi+Pi-}) to obtain the following relation\footnote{To be precise, the definition of axion mixing angles with states that are not mass eigenstates, such as $\eta_{_{ud}}$ and $\eta_s$, goes as follows. Consider the interactions in the (canonically normalized) quark flavor basis:
\be
V~\supset~  \frac{\mu_{q}^2}{2}\,\eta^2_{_{ud}} \,+\, \frac{\mu_s^2}{2}\,\eta^2_s \,+\, \mu^2_{qs}\, \eta_{_{ud}}\eta_s \,+\, \mu_{aq}^2\, a\,\eta_{_{ud}} \,+\, \mu_{as}^2\, a\,  \eta_s \,+\, \mathcal{J}_{_{ud}}\eta_{_{ud}} \,+\, \mathcal{J}_s\eta_s \,,
\ee
where $\mathcal{J}_{_{ud}}$ and $\mathcal{J}_s$ are external sources. Integrating $\eta_{_{ud}}$ and $\eta_s$ out, one obtains:
\be
V~\supset~ \theta_{a\eta_{_{ud}}}a\, \mathcal{J}_{_{ud}}  \,+\,  \theta_{a\eta_s} a\,\mathcal{J}_s \,,
\ee
with
\be
\theta_{a\eta_{_{ud}}}\,=\,-\frac{\mu^2_{aq}\,\mu^2_{s}-\mu^2_{qs}\,\mu^2_{as}}{\mu^2_{q}\,\mu^2_{s}-\mu^4_{qs}}~~~~~,~~~~~
\theta_{a\eta_s}\,=\,-\frac{\mu^2_{as}\,\mu^2_{q}-\mu^2_{qs}\,\mu^2_{aq}}{\mu^2_{q}\,\mu^2_{s}-\mu^4_{qs}}\,.
\ee}:
\be\label{AmpKpiaOctet}
\mathcal{M}(K^+\rightarrow\pi^+ a)\big|_\text{octet enh.}&\approx&\theta_{a\eta_{_{ud}}}\,\sqrt{2}\;\mathcal{M}(K^0\rightarrow\pi^+\pi^-)\nonumber\\
&\approx&\theta_{a\eta_{_{ud}}}\;\mathcal{M}(K_S^0\rightarrow\pi^+\pi^-)\,.
\eeq

Ref.~\cite{Antoniadis:1981zw} noted that expression (\ref{AmpKpiaOctet}) does not take into account the absence of strong final-state interactions between $\pi^+$ and $a$. Following \cite{Antoniadis:1981zw}, we correct this by introducing a fudge factor $D_{\pi\pi}\sim 1/\sqrt{3}$ multiplying the r.h.s. of (\ref{AmpKpiaOctet}). We then finally obtain:
\beq\label{BrKpiaOctet}
\Br(K^+\rightarrow\pi^+ a)\big|_\text{octet enh.}&\approx&\theta_{a\eta_{_{ud}}}^2\,\frac{\Gamma_{K^0_s}}{\;\Gamma_{K^+}}\;\Br(K^0_s\rightarrow\pi^+\pi^-)\;\frac{|\vec{p}_a|}{|\vec{p}_{\pi}|}\;D_{\pi\pi}^2\nonumber\\
&\approx&36\;\theta_{a\eta_{_{ud}}}^2\,.
\eeq
Finally, using the experimental upper bound (\ref{Kexpbound}), we infer the following constraint:
\be\label{OctetLimit}
|\,\theta_{a\eta_{_{ud}}}|\bigm|_\text{octet enh.}~\lsim~(1.7-5.3)\times10^{-4}.
\ee


We would like to stress that the arguments above, leading to (\ref{BrKpiaOctet}) and (\ref{OctetLimit}), hinge on the assumption that the 
$\mathcal{O}(p^2)$ octet operator, shown in (\ref{Loctet27plet}), is enhanced. While this assumption is not considered controversial, it is nonetheless possible that this is not the correct description of octet enhancement in non-leptonic kaon decays \cite{Gerard:2000jj,Buras:2014maa,Crewther:2013vea}. 
In particular, an equally good description is obtained by transfering the enhancement to the $\mathcal{O}(p^4)$ expansion of $\mathcal{L}_{\chi}^{\Delta S=1}$.
Consider, for instance, 
\be\label{LoctetP4}
\mathcal{L}_\chi^{\Delta S=1}\Big|_{\mathcal{O}(p^4)}\supset~g_{8}^\prime\;\frac{f_\pi^2}{\Lambda^2}\,B\;\Tr\left(\lambda_{ds}\, M_q \,U\right)\;\Tr\left(\partial_\mu U\, \partial^\mu U^\dagger\right)\;+\;\text{h.c.}\,,
\ee
where $\Lambda\sim4\pi f_\pi$ is a natural cut-off. If one assumes that the $\mathcal{O}(p^2)$ coefficients $g_8$ and $g_{27}$ are comparable, and instead the $\mathcal{O}(p^4)$ coefficient $g_8^\prime$ is enhanced,
\be\label{g8p4}
\frac{g_8}{g_{27}}\sim\mathcal{O}(1)\,,\qquad 
\frac{g_8^\prime}{g_{27}}\sim\mathcal{O}(100)\,,\qquad 
g_8^\prime\simeq\, \frac{\Lambda^2}{2\,m_K^2} \;g_8\,|_{(\ref{g8g27})}\approx\,1.6\times10^{-7},
\ee
one obtains an equally good phenomenological fit of $K\rightarrow \pi\pi$ and $K\rightarrow\pi\pi\pi$ data compared to the standard fit in (\ref{g8g27}).\footnote{Specifically, there is a degeneracy in the \{$g_8$, $g_8^\prime$\} dependence of the amplitudes for $K\rightarrow \pi\pi$ and $K\rightarrow\pi\pi\pi$, which can only be broken with additional assumptions. See \cite{Kambor:1991ah,Bijnens:2002vr}.}

However, unlike the octet operator in (\ref{Loctet27plet}), (\ref{LoctetP4}) {\it does not contribute} to $K^+\rightarrow\pi^+\eta_{8,0}$. Therefore, the implication of the alternative fit in (\ref{g8p4}) would be that the amplitude for $K^+\rightarrow\pi^+a$, which is proportional to $g_8$, but unaffected by $g_8^\prime$, would not be octet enhanced. In that case, one would have to properly redo the fit to kaon data to obtain the precise value of $g_8$, but as a rough estimate, taking $g_8\sim\mathcal{O}(g_{27})$, the constraint on $a-\eta$ mixing would be weakened by a factor of $\sim30$,
\be\label{NoOctetLimit}
|\,\theta_{a\eta_{_{ud}}}|\bigm|_\text{not octet enh.}~\lsim~(0.5-1.6)\times10^{-2}.
\ee

In short, the enhancement of the amplitude $\mathcal{M}(K^+\rightarrow\pi^+a)$ depends on how octet enhancement is realized in $\chi$PT. If the amplitudes $\mathcal{M}(K^+\rightarrow\pi^+\eta_{8,0})$ are not octet enhanced, the resulting bound (\ref{NoOctetLimit}) on $\theta_{a\eta_{_{ud}}}$ is relatively weak and does not presently pose a threat to the viability of MeV axions. Under the more conventional assumption of octet enhancement at $\mathcal{O}(p^2)$, however, it becomes important to establish how compatible is $\theta_{a\eta_{_{ud}}}$ with the bound in (\ref{OctetLimit}), which we shall now address.

\subsection{Axion-Eta mixing}
\label{aetamix}

The last sources of uncertainty in predicting the rare decay $K^+\rightarrow\pi^+ a$, namely, the mixing angles $\theta_{a\eta_8}$ and $\theta_{a\eta_0}$, have been the least critically examined in the axion literature.

It is well known that the leading order expansion in $\chi$PT does not adequately describe the $\eta$ and $\eta^\prime$ masses and mixing angles \cite{Georgi:1993jn,Gerard:2004gx}. Indeed, the second order expansion of the Chiral Lagrangian provides important corrections to masses, decay constants and mixing angles of singlet and octet mesons, which typically scale as:
\be\label{LECsize}
\frac{32\, m_K^2}{f_\pi^2}\, L_i~\sim~ \mathcal{O}(10^3) \, L_i \, .
\ee
Above, $L_i$ are the dimensionless coefficients of the $\mathcal{O}(p^4)$ operators in the chiral expansion, and are commonly known as {\it Low Energy Constants} (LECs) \cite{Gasser:1983yg,Gasser:1984gg}. Many LECs are reasonably well-determined from experimental and/or lattice data, their typical size being $L_i\sim\mathcal{O}(10^{-3})$. From (\ref{LECsize}) it is then evident that these encode $\mathcal{O}(1)$ effects in $\eta$-$\eta^\prime$ mixing, and may very well have comparable importance in describing $a$-$\eta$ and $a$-$\eta^\prime$ mixing.

In order to illustrate the uncertainties involved in obtaining $\theta_{a\,\eta,\eta^\prime}$, we consider Leutwyler's study of $\eta$-$\eta^\prime$ mixing in \cite{Leutwyler:1997yr}, which, based on $1/N_c$-expansion counting rules, retained only the following operators\footnote{For simplicity, in this exercise we omit the OZI violating terms in (\ref{LeutwylerL}). Their numerical values obtained in the fit of \cite{Leutwyler:1997yr} change our results by $\mathcal{O}(10\%)$.  }:
\beq\label{LeutwylerL}
\mathcal{L}_{\chi}^{\eta-\eta^\prime}&=&\frac{F^2}{4} \, \Tr\left[\, \partial_\mu U\,\partial^\mu U^\dagger\right]\;+\;
\frac{F^2}{4} \, \Tr\left[\, (2BM_q)\, U +\text{h.c.} \right]\;-\;
\frac{1}{2}M_0^2\eta_0^2\\
&+&L_5\,\Tr\left[\, \partial^\mu(2BM_q\,U )\,\partial_\mu U^\dagger\, U +\text{h.c.}\,\right]
 \;+\; L_8\, \Tr\left[\,(2BM_q)\,U\,(2BM_q)\,U +\text{h.c.}\,\right]\nonumber\\
 &+&\text{OZI violating terms.}\nonumber
\eeq
Loop corrections do not contribute at this order, and \cite{Leutwyler:1997yr} obtains $F=90.6$~MeV, $L_5=2.2\times 10^{-3}$, $L_8=1.0\times10^{-3}$, and $M_0\simeq 1030$~MeV.  

Remembering that the axion is formally a phase of the light quark mass matrix $M_q$ (see (\ref{Mqa})), and setting $Q_d=Q_u/2=1$, we can expand (\ref{LeutwylerL}) to obtain:
%
\beq\label{LeutwylerLa}
\mathcal{L}_{\chi}^{\eta-\eta^\prime}&\supset&\frac{1}{2}\partial_\mu\eta_{_{ud}}\partial^\mu\eta_{_{ud}} 
\;+\; \frac{1}{2}\Big(1+\frac{\hat{L}_5}{2}\Big)\partial_\mu\eta_{s}\partial^\mu\eta_{s}  
\;+\; \frac{1}{2}\partial_\mu a\,\partial^\mu a
\nonumber\\
&-& \frac{1}{2}\,2\,(1+\hat{L}_8)\,m_K^2\, \eta_s^2
\;-\;\frac{1}{2}\,M_0^2\,\bigg(\frac{\eta_s}{\sqrt{3}}+\frac{\eta_{_{ud}}}{\sqrt{3/2}}\bigg)^2
\;-\; \frac{1}{2}\,m_a^2\,a^2\nonumber
\\
&+& \frac{\hat{L}_5}{6}\,\frac{m_\pi^2}{m_K^2}\,\frac{F}{f_a}\,\partial_\mu\eta_{_{ud}}\partial^\mu a
\;-\;  \frac{4\,m_{\pi}^2}{3}\,\frac{F}{f_a}\;\eta_{_{ud}}a\,.\nonumber
\eeq
Above, we have expanded the coefficients in powers of $\frac{m_\pi^2}{m_K^2}$ and kept only the leading terms; we have also omitted all $\pi$ and $K$ dependent terms, since they are not relevant for the present discussion. Furthermore, we have simplified the notation by defining:
\be
\hat{L}_i~\equiv~\frac{32\, m_K^2}{F^2}\, L_i\,.
\ee
Canonically normalizing the kinetic terms and integrating out $\eta_{_{ud}}$ and $\eta_s$, we obtain:
\be
\theta_{a\eta_{_{ud}}}\,=\,-\bigg(   \frac{1-\hat{L}_5(1+\hat{L}_8)/2}{3\,(1+\hat{L}_8)}\,+\,2\,\frac{m_K^2}{M_0^2}    \bigg) \,\frac{m_\pi^2}{m_K^2}\,\frac{F}{f_a}\,.
\ee
Plugging in the numerical fit of the LECs obtained by \cite{Leutwyler:1997yr} yields:
\be\label{aEtaMixLeutNumber}
\theta_{a\eta_{_{ud}}}~~\approx~~\frac{-1.9\times10^{-3}}{f_a/\text{GeV}}~~\approx~~\frac{1}{3}\; \theta_{a\eta_{_{ud}}}^{(0)}\,.
\ee

Rather than claiming that we have calculated $\theta_{a\eta_{_{ud}}}$ more precisely, our purpose with this exercise is to make the point that contributions from the second order chiral expansion change the leading order estimate, $\theta_{a\eta_{_{ud}}}^{(0)}$, by $\mathcal{O}(1)$. While we have restricted this exercise to a couple of LECs at tree level, inclusion of the full $\mathcal{O}(p^4)$ expansion and loop corrections will most likely change the answer from (\ref{aEtaMixLeutNumber}). Had we derived such expression, even then we would not be able to evaluate it numerically because many of the LECs contributing to $\theta_{a\eta_{_{ud}}}$ are still undetermined. For instance, operators such as \cite{Borasoy:2001ik}
\beq
\mathcal{L}_{\chi}^{\mathcal{O}(p^4)}&\supset&L_7\, \Tr\left[\,(2BM_q)\,U-U^\dagger\,(2BM_q)^\dagger \,\right]^2\,+\,i\,\lambda_2\,F^2\,\frac{\eta_0}{F}\,\Tr\left[\,(2BM_q)\,U-U^\dagger\,(2BM_q)^\dagger \,\right]\nonumber\\
&+&L_{18}\, \, \Tr\left[\,U^\dagger\partial^\mu U \,\right]\,\Tr\left[\,\partial_\mu\big(U^\dagger\,(2BM_q)^\dagger- (2BM_q)\,U\big)\,\right] \nonumber\\
&+&i\,L_{25}\,\frac{\eta_0}{F}\, \Tr\left[\,U^\dagger\,(2BM_q)^\dagger\,U^\dagger\,(2BM_q)^\dagger-(2BM_q)\,U\,(2BM_q)\,U\,\right] \\
&+&i\,L_{26}\,\frac{\eta_0}{F}\,\Big( \Tr\left[\,U^\dagger\,(2BM_q)^\dagger\,\right]^2-\Tr\left[\,(2BM_q)\,U\,\right]^2\Big)\,+...\nonumber
\eeq
also give important contributions $\theta_{a\eta_{_{ud}}}$ and $\theta_{a\eta_s}$. While such operators also affect $\eta$-$\eta^\prime$ mixing, and hence are subject to constraints, there are too many independent LECs to be fit by observations, and without further assumptions, the parameterization of the $\eta$-$\eta^\prime$ phenomenology in $\chi$PT is underconstrained.

We conclude, therefore, that currently it is impossible to reliably estimate $\theta_{a\eta_{_{ud}}}$ and $\theta_{a\eta_s}$, given that these mixings angles can receive $\mathcal{O}(1)$ corrections from the second order chiral expansion whose numerical values are undetermined by existing data. Therefore, one cannot claim with confidence that $\theta_{a\eta_{_{ud}}}$ violates the $K^+\rightarrow\pi^+a$ bounds in (\ref{OctetLimit}) or (\ref{NoOctetLimit}), and therefore $K^+$ decay bounds by themselves do not provide a {\it definitive} exclusion of the QCD axion variant we are considering. Most likely, a reliable determination of the mixing angles $\theta_{a\eta_{_{ud}}}$ and $\theta_{a\eta_s}$ would have to come from lattice calculations, or, were such axion to be realized in nature, from direct measurements.

\section{\label{GammaGamma} The Physical Axion Current}

Because of much confusion in the literature, in this section we review the differences between the PQ current, the Bardeen-Tye current, and the current associated with the axion mass eigenstate. Identifying the latter, in particular, is a critical part of working out the proper phenomenology of axions, since it is the physical axion current that determines the coupling of the axion to the chiral $U(3)_\chi$ currents that mediate meson decays, nuclear de-excitations, as well as the electromagnetic anomaly that induces the axion coupling to two photons.

In the UV, the PQ current for the class of axion variants defined in (\ref{Laqq}) is given by\footnote{Since this discussion is concerned with the hadronic properties of the axion, we ignore the leptonic current and electromagnetic anomaly for the remainder of this subsection.}:
\be\label{JPQ}
J_\mu^{\text{PQ}}\; =\; f_a\,\partial_\mu a \,-\, \frac{Q_u}{2}\, \bar{u}\gamma_\mu\gamma_5u\,\,-\, \frac{Q_d}{2}\, \bar{d}\gamma_\mu\gamma_5d \,.
\ee

Bardeen and Tye \cite{Bardeen:1977bd} pointed out that the PQ current cannot be associated with the physical axion, i.e., the light mass eigenstate, since the divergence of this current receives a large contribution from the strong anomaly:
\be\label{divJPQ}
\partial^\mu J_\mu^{\text{PQ}}~=~ -\frac{\alpha_s}{8\pi}\, (Q_u+Q_d)\, G_{\mu\nu}\tilde{G}^{\mu\nu} \,,
\ee
which does not vanish in the limit $m_q\rightarrow 0$. The physical axion, on the other hand, becomes a {\it bona fide} Goldstone boson in the limit of a massless quark, and should therefore be associated with a current that is conserved as $m_q\rightarrow 0$.

In order to identify the physical current, \cite{Bardeen:1977bd} subtracted the strong anomaly from $J^{\text{PQ}}$, obtaining the {\it Bardeen-Tye} current:
\be\label{JBT}
J^{\text{BT}}_\mu~\equiv~ J_\mu^{\text{PQ}}\,+\,\frac{(Q_u+Q_d)/2}{(m_u^{-1}+m_d^{-1}+m_s^{-1})}\,\big(m_u^{-1}\,\bar{u}\gamma_\mu\gamma_5u+m_d^{-1}\,\bar{d}\gamma_\mu\gamma_5d+m_s^{-1}\,\bar{s}\gamma_\mu\gamma_5s\big),
\ee
which is conserved in the limit $m_q\rightarrow 0$:
\be\label{divJBT}
\partial^\mu J_\mu^{\text{BT}}~=~ (Q_u+Q_d)\;\frac{m_um_dm_s}{m_dm_s+m_um_s+m_um_d}\;\big(\bar{u}i\gamma_5u+\bar{d}i\gamma_5d+\bar{s}i\gamma_5s\big)  \,.
\ee

We can rewrite $J^{\text{BT}}$ in order to show its explicit dependence on the chiral $U(3)_\chi$ currents. First we define:
\begin{subequations}\label{Jchi}
\begin{alignat}{3}
J_{5\,\mu}^{\,(3)} &~=~ \frac{\bar{u}\gamma_\mu\gamma_5u - \bar{d}\gamma_\mu\gamma_5d}{2} &&~\equiv~f_\pi\, \partial_\mu\pi_3 \,,
\label{J3}\\
J_{5\,\mu}^{\,(8)} &~=~ \frac{\bar{u}\gamma_\mu\gamma_5u + \bar{d}\gamma_\mu\gamma_5d -2 \bar{s}\gamma_\mu\gamma_5s}{2\sqrt{3}} &&~\equiv~ f_\pi\, \partial_\mu\eta_8 \,,
\label{J8}\\
J_{5\,\mu}^{\,(0)} &~=~ \frac{\bar{u}\gamma_\mu\gamma_5u + \bar{d}\gamma_\mu\gamma_5d + \bar{s}\gamma_\mu\gamma_5s}{\sqrt{6}} &&~\equiv~\, f_\pi\,  \partial_\mu\eta_0 \,.
\label{J0}
\end{alignat}
\end{subequations}
Above, $\pi_3$, $\eta_8$ and $\eta_0$ are {\it not} mass eigenstates.

The Bardeen-Tye current in (\ref{JBT}) can then be written as:
%
\beq\label{JBTrewritten}
\hspace{-0.9cm}
J^{\text{BT}}_\mu&=&f_a\,\partial_\mu a-\bigg(\frac{(Q_um_u-Q_dm_d)}{(1+\epsilon_\eta)(m_u+m_d)} +\epsilon_\eta\,\frac{(Q_u-Q_d)}{2(1+\epsilon_\eta)}  \bigg)\,J_{5\,\mu}^{\,(3)}
-\bigg(\frac{\sqrt{3}}{~2}\frac{\epsilon_\eta}{(1+\epsilon_\eta)}\,(Q_u+Q_d)\!\bigg)\,J_{5\,\mu}^{\,(8)}\,,~~~~~~~
\eeq
\beq
\epsilon_\eta~\equiv~\frac{m_um_d}{m_s(m_u+m_d)}~\approx~\frac{m_um_d}{(m_u+m_d)^2}\,\frac{m_\pi^2}{m_K^2}\,.
\eeq

The Bardeen-Tye current has been identified in numerous studies as the physical axion current. Under this assumption, it is straightforward to read the mixing angles $\theta_{a\,\pi/\eta}$ from (\ref{Jchi}) and (\ref{JBTrewritten}):
\beq
\theta_{a\pi}^{\text{BT}} &=& -\frac{1}{(1+\epsilon_{\eta})}\;\bigg(\frac{(Q_u m_u-Q_d m_d)}{(m_u+m_d)}+\epsilon_{\eta}\,\frac{(Q_u-Q_d)}{2}\bigg)\,\frac{f_\pi}{f_a}\;, \label{theta-a-pi-BT}\\
\theta_{a\eta_8}^{\text{BT}} &=& -\frac{\sqrt{3}}{~2}\;\frac{\epsilon_{\eta}}{(1+\epsilon_{\eta})}\;(Q_u+Q_d)\; \frac{f_\pi}{f_a}\;, \label{theta-a-8-BT}\\
\theta_{a\eta_0}^{\text{BT}} &=&0\;.\label{theta-a-0-BT}
\eeq

Note, however, that this assumption is only strictly correct in the limit where the $\eta^\prime$ is decoupled, i.e., in the limit $M_0\rightarrow\infty$ (see (\ref{chiL0})) and $\eta^\prime\rightarrow\eta_0$. Indeed, the Bardeen-Tye prescription to obtain the mixing angles $\theta_{a\,\pi/\eta}$ is {\it equivalent} to the $1^{\text{st}}$ order $\chi$PT prescription with $\eta_0$ decoupled. This is corroborated by the fact that (\ref{theta-a-pi}, \ref{theta-a-8}, \ref{theta-a-0}) and (\ref{theta-a-pi-BT}, \ref{theta-a-8-BT}, \ref{theta-a-0-BT}) agree in the limit $M_0\rightarrow\infty$.

We point out as a side remark that the condition that $\partial^\mu J_\mu^{\text{BT}}$ vanishes in the limit $m_q\rightarrow 0$ is not sufficient to uniquely determine the Bardeen-Tye current. In particular, one could define:
\be\label{JBTprime}
J^{\,\prime\,\text{BT}}_\mu~=~J_\mu^{\text{BT}}+\frac{m_um_dm_s}{(m_u+m_d)(m_u+m_s)(m_d+m_s)}\,\big(c_3\, J_{5\,\mu}^{\,(3)} \,+\, c_8\, J_{5\,\mu}^{\,(8)} \,+\, c_0\, J_{5\,\mu}^{\,(0)}\big),~~~
\ee
which also satisfies the aforementioned condition with arbitrary finite coefficients $c_{3,8,0}$. Again, in this context, this ambiguity can only be removed in the limit of decoupled $\eta^\prime$. In this limit, $c_0$ in (\ref{JBTprime}) must necessarily vanish, since $\theta_{a\eta_0}\propto c_0\rightarrow 0$ as $m_{\eta^\prime}\rightarrow\infty$.\footnote{The vanishing of $\theta_{a\eta_0}$ as $M_0\rightarrow\infty$ only holds for bases where the coupling $(a/f_a)\,G\tilde{G}$ has been rotated away, which is the case of (\ref{Laqq}). However, the statement that the Bardeen-Tye prescription is equivalent to taking the limit $M_0\rightarrow\infty$ in $1^{\text{st}}$ order $\chi$PT is basis-independent.} Moreover, in this limit the mass eigenstates $a_\text{phys}$, $\pi$ and $\eta$ are entirely determined by the non-anomalous currents $J^{\,\prime\,\text{BT}}$, $J^{(3)}$ and $J^{(8)}$. Imposing then that $J^{\,\prime\,\text{BT}}$ commutes with $J^\pi$ and $J^\eta$, one can estimate $c_3$ and $c_8$ via current algebra methods, and conclude that these coefficients provide only small corrections to (\ref{theta-a-pi-BT}) and (\ref{theta-a-8-BT}) of order $m_a^2/m_\pi^2$ and $m_a^2/m_\eta^2$, respectively.

Unfortunately, once the $\eta^\prime$ is included in the spectrum, the Bardeen-Tye prescription is no longer sufficient to determine the physical axion current. In particular, while $\pi_3$ in (\ref{J3}) is a good approximation to the mass eigenstate $\pi^0$, the physical states $\eta$ and $\eta^\prime$ have substantial components of both $\eta_8$ and $\eta_0$, and hence are not amenable to current algebra methods, which are not applicable to anomalous currents.

Therefore, generically the physical axion current has components from all three neutral chiral $U(3)_\chi$ currents:
\be\label{Jaxion}
J_\mu^{\,a_\text{phys}}\;\equiv\;  f_a \partial_\mu a_\text{phys}\; =\; f_a\,\partial_\mu a \,+\, \xi_{a3}\, J_{5\,\mu}^{\,(3)} \,+\, \xi_{a8}\, J_{5\,\mu}^{\,(8)} \,+\, \xi_{a0}\, J_{5\,\mu}^{\,(0)} \,.
\ee

The coefficients $\xi_{a3}$, $\xi_{a8}$ and $\xi_{a0}$ in (\ref{Jaxion}) are related to the axion-meson mixing angles by:
\be\label{mixings}
\theta_{a\pi}=\xi_{a3}\frac{f_\pi}{f_a}\quad,\quad
\theta_{a\eta_8}=\xi_{a8}\frac{f_\pi}{f_a}\quad,\quad
\theta_{a\eta_0}=\xi_{a0}\frac{f_\pi}{f_a}.
\ee

This discussion provides yet another perspective on the difficulties in estimating the hadronic mixing angles of the axion. This propagates into uncertainties in the axion's phenomenology, such as rates of rare meson decays, discussed in the previous two sections, as well as the axion couplings to photons and nucleons, which we will now consider.

\subsection{Axion-photon coupling}
\label{axion-photon}

As light quarks confine below the QCD scale, an effective operator coupling the axion to the electromagnetic dual field strength is generated. In this section we derive this operator, estimate the axion decay branching ratio to a pair of photons, and comment on bounds from $K^+\rightarrow\pi^+(a\rightarrow\gamma\gamma)$.

We start by rewriting the physical axion current (\ref{Jaxion}) as:
\be\label{JaxionQbasis}
J_\mu^{\,a_\text{phys}}\; =\; f_a\,\partial_\mu a \,+\,\xi_{u}\,\bar{u}\gamma_\mu\gamma_5u+\xi_{d}\,\bar{d}\gamma_\mu\gamma_5d+\xi_{s}\,\bar{s}\gamma_\mu\gamma_5s\,,
\ee
where
\be\label{mixings}
\xi_{u}&\;=\;&\frac{f_a}{f_\pi}\,\bigg(\frac{\theta_{a\pi}}{2}+\frac{\theta_{a\eta_8}}{2\sqrt{3}}+\frac{\theta_{a\eta_0}}{\sqrt{6}}\bigg)
\;=\;\frac{f_a}{f_\pi}\,\bigg(\frac{\theta_{a\pi}}{2}+\frac{\theta_{a\eta_{_{ud}}}}{2}\bigg)\,,\\
\xi_{d}&\;=\;&\frac{f_a}{f_\pi}\,\bigg(\!\!-\frac{\theta_{a\pi}}{2}+\frac{\theta_{a\eta_8}}{2\sqrt{3}}+\frac{\theta_{a\eta_0}}{\sqrt{6}}\bigg)
\;=\;\frac{f_a}{f_\pi}\,\bigg(\!\!-\frac{\theta_{a\pi}}{2}+\frac{\theta_{a\eta_{_{ud}}}}{2}\bigg)\,\\
\xi_{s}&\;=\;&\frac{f_a}{f_\pi}\,\bigg(\!\!-\,\frac{\theta_{a\eta_8}}{\sqrt{3}}+\frac{\theta_{a\eta_0}}{\sqrt{6}}\bigg)
\;=\;\frac{f_a}{f_\pi}\,\bigg(\frac{\theta_{a\eta_{_{s}}}}{\sqrt{2}}\bigg)\,.
\ee

With this notation, we can derive the electromagnetic anomaly of $J_\mu^{\,a_\text{phys}}$ and the axion coupling to photons straightforwardly:
\beq\label{aFFdual}
\mathcal{L}_a\;&\supset&\;\frac{\alpha}{2\pi f_a}\,3\left(\xi_u\left(\frac{2}{3}\right)^2+\xi_d\left(\!-\frac{1}{3}\right)^2+\xi_s\left(\!-\frac{1}{3}\right)^2\right)\,a\; F_{\mu\nu}\tilde{F}^{\mu\nu}\nonumber\\
&=&\;\frac{\alpha}{4\pi f_\pi}\,\left(\theta_{a\pi}+\frac{\theta_{a\eta_8}}{\sqrt{3}}+2\,\frac{\theta_{a\eta_0}}{\sqrt{3/2}}\right)\,a\; F_{\mu\nu}\tilde{F}^{\mu\nu}\nonumber\\
&=&\;\frac{\alpha}{4\pi f_\pi}\,\left(\theta_{a\pi}+\frac{5}{3}\,\theta_{a\eta_{_{ud}}}+\frac{\sqrt{2}}{~3}\,\theta_{a\eta_{_{s}}}\right)\,a\; F_{\mu\nu}\tilde{F}^{\mu\nu}\,,
\eeq
where we assume that $m_a> 2 \,m_e$, so that the coupling of the axion to electrons does not contribute to this effective operator.

The axion decay width to $\gamma\gamma$ is then given by:
\be\label{aggWidth}
\Gamma(a\rightarrow\gamma\gamma)\;=\;\left(\theta_{a\pi}+\frac{5}{3}\,\theta_{a\eta_{_{ud}}}+\frac{\sqrt{2}}{~3}\,\theta_{a\eta_{_{s}}}\right)^2\left(\frac{\alpha}{4\pi f_\pi}\right)^2\,\frac{m_a^3}{4\pi}\,.
\ee
This expression neglects the loop induced contribution coming from the axion coupling to electrons, which becomes comparable to the hadronic contribution above if the linear combination of mixing angles in (\ref{aggWidth}) is $\mathcal{O}(10^{-4})$.

The relative $\gamma\gamma$ to $e^+e^-$ branching ratio is then:
\beq\label{BRagg}
\frac{\text{Br}(a\rightarrow\gamma\gamma)}{\text{Br}(a\rightarrow e^+e^-)}=
\frac{2}{\beta_e}\frac{m_u m_d}{(m_u+m_d)^2}\bigg(\frac{Q_u+Q_d}{Q_e}\bigg)^2\bigg(\frac{\alpha}{4\pi}\bigg)^2\bigg(\frac{m_\pi}{m_e}\bigg)^2\left(\theta_{a\pi}+\frac{5}{3}\,\theta_{a\eta_{_{ud}}}+\frac{\sqrt{2}}{~3}\,\theta_{a\eta_{_{s}}}\right)^2~~~~~~
\eeq
where $\beta_e=\sqrt{1-4\,m_e^2/m_a^2}$. Note the very weak dependence of this branching ratio on the axion mass when $m_a\gg m_e$.

From (\ref{BRagg}) we can estimate the branching ratio for the rare decay $K^+\rightarrow\pi^+(a\rightarrow\gamma\gamma)$ in terms $\text{Br}(K^+\rightarrow\pi^+(a\rightarrow e^+e^-))$: 
\beq
\text{Br}(K^+\rightarrow\pi^+(a\rightarrow\gamma\gamma))\approx
\frac{10^{-11}}{Q_e^2}\left(\frac{\theta_{a\pi}+\frac{5}{3}\,\theta_{a\eta_{_{ud}}}+\frac{\sqrt{2}}{~3}\,\theta_{a\eta_{_{s}}}}{10^{-2}}\right)^2\left(\frac{\text{Br}(K^+\rightarrow\pi^+(a\rightarrow e^+e^-))}{10^{-6}}\right),~~~~~~~
\eeq
which is still below current experimental sensitivity by two or three orders of magnitude \cite{Artamonov:2005ru}.

\subsection{Axion-nucleon couplings and nuclear de-excitations}
\label{axion-nucleon}

Another classic signature of visible axions is axion emission in nuclear de-excitations, first studied in \cite{Treiman:1978ge, Donnelly:1978ty, Barroso:1981bp}. We shall briefly review here the standard results from the literature.

From (\ref{Jaxion}) or (\ref{JaxionQbasis}) we can infer the axion coupling to protons and neutrons:
\beq
\mathcal{L}_{aNN}= \overline{N}\,i\gamma_5\;g_{aNN}\;a\;N\,,
\eeq
\beq\label{gaNN}
g_{aNN} = \frac{2m_N}{f_\pi}\,\bigg(\frac{\theta_{a\pi}}{2}\,g_A^{(3)}\,\tau^3\;+\;\frac{\theta_{a\eta_8}}{2\sqrt{3}}\;g_A^{(8)}\;+\;\frac{\theta_{a\eta_0}}{\sqrt{6}}\;g_A^{(0)}\bigg)\,,
\eeq
where $N=\binom{\,p\,}{\,n\,}$ is the nucleon isospin doublet, $m_N$ is the nucleon mass, and $\tau^3=\bigl(\begin{smallmatrix} 1& \\ &\!\!-1\! \end{smallmatrix} \bigr)$ is a Pauli matrix. The axial coupling constants, $g_A^{(i)}$, are the axial-vector form factors of $\langle N| J_5^i\,|N \rangle$ at $q^2=0$,
\beq
g_A^{(3)}~&=&~\Delta u - \Delta d\,,\\
g_A^{(8)}~&=&~\Delta u + \Delta d -2\,\Delta s\,,\\
g_A^{(0)}~&=&~\Delta u + \Delta d + \Delta s\,,
\eeq
where $\Delta q$ is defined as:
\be
2\,s_\mu\,\Delta q = \langle N| \bar q\gamma_\mu\gamma_5 q\,|N \rangle\,
\ee
with $s_\mu$ being the nucleon spin-vector. In terms of $\Delta q$, we can alternatively recast $g_{aNN}$ as:
\beq\label{gaNNDeltaq}
g_{aNN} = \frac{m_N}{f_\pi}\,\Big(\theta_{a\pi}\,(\Delta u - \Delta d)\,\tau^3\;+\;\theta_{a\eta_{_{ud}}}(\Delta u + \Delta d) \;+\; \sqrt{2}\,\theta_{a\eta_{_{s}}}\Delta s\,\Big).
\eeq

The pion-nucleon coupling, set by $g_A^{(3)}$, is well determined from nuclear $\beta$-decay,
\be
g_A^{(3)} \; =\; \Delta u - \Delta d\;\simeq\; 1.27\,.
\ee
The other axial coupling constants, $g_A^{(8)}$ and $g_A^{(0)}$, are much harder to extract from experiment, and various estimates based on data from semi-leptonic hyperon decays, proton deep inelastic scattering, and lattice calculations yield the following ranges \cite{Jaffe:1989jz, Savage:1996zd, Karliner:1999fn, Mallot:1999qb, Leader:2000dw, Cheng:2012qr, QCDSF:2011aa, Engelhardt:2012gd, Abdel-Rehim:2013wlz, Bhattacharya:2015gma, Abdel-Rehim:2015owa, Abdel-Rehim:2015lha, Green:2017keo}:
\beq
0.09\lesssim~\Delta u + \Delta d~\lesssim0.62\qquad \text{and}\qquad -0.35\lesssim \Delta s\lesssim 0.
\eeq
These uncertainties, compounded by the difficulties in extracting the axion hadronic mixing angles $\theta_{a\,\pi/\eta}$, are obstacles to making firm predictions for nuclear de-excitation rates via axion
emission\footnote{Ref.\cite{diCortona:2015ldu} also considered properties of the axion in great detail, extracting the axion mass and coupling to photons and nucleons by combining results from Lattice QCD (LQCD) and $\chi$PT at NLO. We believe that the authors of \cite{diCortona:2015ldu} were overly optimistic regarding the smallness of the errors in some of the LECs used, as well as the errors in LQCD extractions of the nucleon spin content $\Delta q$. But more importantly, their approach differs substantially from ours. For instance, they associate the axion with the Bardeen-Tye current in the two-flavor effective theory (i.e., chiral EFT with the strange quark integrated out). While the inclusion of NLO contributions in $\chi$PT should in principle correct for the use of the (non physical) Bardeen-Tye current, it is unclear to us whether the authors of \cite{diCortona:2015ldu} did it in a consistent manner such that the resulting axion couplings to photons and nucleons would indeed correspond to that of the mass eigenstate. For instance, applying their procedure to our model to extract the couplings of the axion to nucleons, one would be led to conclude that our axion variant does not couple to nucleons, which is incorrect. (In particular, in their notation, our model would correspond to: $c_u^0=Q_u/3$, $c_d^0=Q_d/3$, $c_{s,c,b,t}^0=0$, $f_a \mapsto (Q_u+Q_d)/f_a$, $E/N=2$ in eq.\,(1) of \cite{diCortona:2015ldu}; this would translate into $c_u=c_d=c_{s,c,b,t}=0$ in eq.\,(46) of \cite{diCortona:2015ldu}). A thorough comparison of our treatment with that of \cite{diCortona:2015ldu} is a non-trivial task which is beyond the scope of this paper.}.

Nevertheless, the parametric dependence of such rates on nuclear and axion properties is relatively well understood. Donnelly et al.~\cite{Donnelly:1978ty} showed that the axion acts as a ``magnetic photon'' in nuclear transitions, and using standard multipole techniques in the long-wavelength limit, derived the ratio of axion to photon de-excitation rates of a nuclear state. For an isoscalar ($\Delta T=0$), $p$-wave magnetic (M1) transition, \cite{Donnelly:1978ty} derived:
\beq
\frac{\Gamma_a}{\Gamma_\gamma}\;\Bigg|_{\Delta T=0}&=&~\frac{1}{2\pi\alpha}\left(\frac{g_{aNN}^{(0)}}{\mu^{(0)}-\eta^{(0)}}\right)^2\left(\frac{|\vec{p}_a|}{|\vec{p}_\gamma|}\right)^{3}\\
&\approx&~\frac{1}{2\pi\alpha}\frac{m_N^2}{f_\pi^2}\left(\frac{\theta_{a\eta_{_{ud}}}(\Delta u + \Delta d) + \sqrt{2}\,\theta_{a\eta_{_{s}}}\Delta s}{\mu^{(0)}-\eta^{(0)}}\right)^2\left(1-\frac{m_a^2}{{\Delta E}^{\,2}}\right)^{3/2}.
\eeq
Above, $\Delta E$ is the energy splitting between the two nuclear levels. The parameters $\mu^{(0)}\approx 0.88$ and $\eta^{(0)}\approx 1/2$ are related to the nuclear magnetic moment and the ratio of convection to magnetization currents, respectively.

For an isovector ($\Delta T=1$) M1 transition, the analogous expression is \cite{Donnelly:1978ty}:
\beq
\frac{\Gamma_a}{\Gamma_\gamma}\;\Bigg|_{\Delta T=1}&=&~\frac{1}{2\pi\alpha}\left(\frac{g_{aNN}^{(1)}}{\mu^{(1)}-\eta^{(1)}}\right)^2\left(\frac{|\vec{p}_a|}{|\vec{p}_\gamma|}\right)^{3}\\
&\approx&~\frac{1}{2\pi\alpha}\left(\frac{m_N\,g_A^{(3)}}{f_\pi}\right)^2\left(\frac{\theta_{a\pi}}{\mu^{(1)}}\right)^2\left(1-\frac{m_a^2}{{\Delta E}^{\,2}}\right)^{3/2},
\eeq
where $\left(\mu^{(1)}-\eta^{(1)}\right)\approx\mu^{(1)}\approx4.7$. For mixed isospin transitions, the generalization is \cite{Savage:1988rg}:
\beq\label{nuclearRateMixed}
\frac{\Gamma_a}{\Gamma_\gamma}
\;&=&\;\frac{1}{2\pi\alpha}\left(\frac{c_0\, g_{aNN}^{(0)}+c_1\, g_{aNN}^{(1)}}{c_0\,(\mu^{(0)}-\eta^{(0)})+c_1\,(\mu^{(1)}-\eta^{(1)})}\right)^2\left(\frac{|\vec{p}_a|}{|\vec{p}_\gamma|}\right)^{3},
\eeq
where $c_{0,1}$ are the probability amplitudes for the different isospin components.

During the 80's, several experiments have searched for variants of the visible QCD axion in nuclear de-excitations \cite{Calaprice:1979pe, Savage:1986ty, Hallin:1986gh, DeBoer:1986cm, Hallin:1987uc, Savage:1988rg, Datar:1988ju, Freedman:1989gz, Asanuma:1990rm}. Since most nuclear levels have splittings of a few MeV, the results of these experiments were only relevant for a range of $m_a$ already ruled-out by other measurements, in particular beam dump experiments (see Sec.~\ref{Sec:AxionElectronCoupling}). In 1990, de Boer et al. \cite{deBoer:1990yg} studied transitions of the $^8$Be nucleus which were energetic enough to probe axion masses of up to $\approx 15$ MeV. A few years later, however, the results of this experiment were revisited by the authors in \cite{deBoer:1996qdk, deBoer:1997glc} and the claimed limits were weakened substantially. We shall therefore not consider the results of \cite{deBoer:1990yg, deBoer:1996qdk, deBoer:1997glc} here.

Most recently, the ATOMKI collaboration \cite{Krasznahorkay:2015iga} has measured several $^8$Be nuclear transitions
via emission of $e^+e^-$ pairs. The two relevant transitions for our discussion are the M1 de-excitations of the $J^P=1^+$ isospin doublet states, namely, $^{\,8}\text{Be}^{*}(17.64)$ and $^{\,8}\text{Be}^{*}(18.15)$, to the $J^P=0^+$ isospin singlet ground state, $^{\,8}\text{Be}(0)$:
\beq
^{\,8}\text{Be}^{*}(17.64)&\rightarrow& \,^{\,8}\text{Be}(0)~~,~~\Delta E=17.64~\text{MeV}~~,~~\Delta T\approx1\,,\\
^{\,8}\text{Be}^{*}(18.15)&\rightarrow& \,^{\,8}\text{Be}(0)~~,~~\Delta E=18.15~\text{MeV}~~,~~\Delta T\approx0\,.
\eeq
 
 The ATOMKI collaboration claimed to have observed a deviation in the $e^+e^-$ spectrum of the $^{\,8}\text{Be}^{*}(18.15)\rightarrow \,^{\,8}\text{Be}(0)$ transition relative to the SM prediction of
internal pair conversion ($\gamma^*\rightarrow e^+e^-$). According to \cite{Krasznahorkay:2015iga}, this deviation was consistent with the on-shell emission of a narrow resonance $X$ of mass $m_X\approx (16.6\pm0.9)$~MeV promptly decaying to $e^+e^-$. The best fit for the relative de-excitation rate was $\Gamma_X/\Gamma_\gamma\approx 5.8\times10^{-6}$, with a statistical significance of $6.8\,\sigma$. Moreover, in the original publication \cite{Krasznahorkay:2015iga}, no excess was observed in the $e^+e^-$ spectrum of the $^{\,8}\text{Be}^{*}(17.64)\rightarrow \,^{\,8}\text{Be}(0)$ transition. No error bars were quoted for either measurement, neither were upper bounds on emission rates of generic new particles, such as light vectors or pseudoscalars. Subsequent studies have attempted to understand these results via nuclear physics models \cite{Zhang:2017zap}, or via emission of a new light resonance \cite{Feng:2016jff,Gu:2016ege,Feng:2016ysn,Ellwanger:2016wfe,Kahn:2016vjr,Fayet:2016nyc,Kozaczuk:2016nma,DelleRose:2017xil}.

The ATOMKI collaboration is revisiting these measurements with an improved experimental set-up \cite{Krasznahorkay:2017qfd, Krasznahorkay:2017gwn}. At the time of writing of this paper, preliminary results by Krasznahorkay et al. have been released \cite{Krasznahorkay:2017gwn} indicating a possible excess in the $^{\,8}\text{Be}^{*}(17.64)\rightarrow \,^{\,8}\text{Be}(0)$ transition as well, which was fit by a similar hypothetical particle $X$ with mass $m_X=(17.0\pm0.2)$~MeV, at a rate of $\Gamma_X/\Gamma_\gamma\approx 4.0\times10^{-6}$.

At present, we believe the claims from the ATOMKI collaboration are inconclusive, and an independent measurement is warranted. Nonetheless, we can make order of magnitude predictions for axion emission in $^8$Be M1 transitions using (\ref{nuclearRateMixed}). While the excited states $^{\,8}\text{Be}^{*}(17.64)$ and $^{\,8}\text{Be}^{*}(18.15)$ are predominantly $T=1$ and $T=0$, respectively, their widths strongly indicate that they are isospin-mixed:
\beq
|^{\,8}\text{Be}^{*}(17.64)\,\rangle&=&\sin\theta_{1^+}\,|\,T=0\,\rangle \,+\, \cos\theta_{1^+}\,|\,T=1\,\rangle,\\
|^{\,8}\text{Be}^{*}(18.15)\,\rangle&=&\cos\theta_{1^+}\,|\,T=0\,\rangle \,-\, \sin\theta_{1^+}\,|\,T=1\,\rangle.
\eeq
Estimates on the level of isospin mixing vary between $0.18\lesssim \sin\theta_{1^+}\lesssim 0.43$ \cite{Pastore:2014oda, Feng:2016ysn, Kozaczuk:2016nma}. Taking the value of $\sin\theta_{1^+}=0.35$ suggested in \cite{Kozaczuk:2016nma}, we obtain the following range for the $^{\,8}\text{Be}^{*}(18.15)$ axion de-excitation rate:
\beq\label{IsoscalarRate8Be}
\frac{\Gamma_a}{\Gamma_\gamma}\;\Bigg|_{^{\,8}\text{Be}^{*}(18.15)}\!\!\!
&=&\frac{1}{2\pi\alpha}\left(\frac{\cos\theta_{1^+}\, g_{aNN}^{(0)}-\sin\theta_{1^+}\, g_{aNN}^{(1)}}{\cos\theta_{1^+}\,(\mu^{(0)}-\eta^{(0)})- \sin\theta_{1^+}\,(\mu^{(1)}-\eta^{(1)})}\right)^2\left(1-\frac{m_a^2}{(18.15\,\text{MeV})^2}\right)^{3/2}\nonumber\\
&&\nonumber\\
&\approx&(1-11)\times10^{-6}~~~~~\text{for}~~~~~\frac{\big|\theta_{a\eta_{_{ud}}}(\Delta u + \Delta d) + \sqrt{2}\,\theta_{a\eta_{_{s}}}\Delta s\big|}{10^{-4}}~\sim~(1.4-4).~~~~~~~~~~~~
\eeq
Above, we also assume $m_a=16.6$ MeV and $\theta_{a\pi}=0.5\times10^{-4}$. Reiterating our point, while uncertainties in nuclear and axion parameters preclude us from making a precise prediction for the rate of axion emission in the de-excitation of $^{\,8}\text{Be}^{*}(18.15)$, the range in (\ref{IsoscalarRate8Be}) is compatible with the excess observed by the ATOMKI collaboration.

Analogously, for the $^{\,8}\text{Be}^{*}(17.64)$ de-excitation we obtain:
\beq\label{IsovectorRate8Be}
\frac{\Gamma_a}{\Gamma_\gamma}\;\Bigg|_{^{\,8}\text{Be}^{*}(17.64)}\!\!\!
&=&\frac{1}{2\pi\alpha}\left(\frac{\sin\theta_{1^+}\, g_{aNN}^{(0)}+\cos\theta_{1^+}\, g_{aNN}^{(1)}}{\sin\theta_{1^+}\,(\mu^{(0)}-\eta^{(0)})+ \cos\theta_{1^+}\,(\mu^{(1)}-\eta^{(1)})}\right)^2\left(1-\frac{m_a^2}{(17.64\,\text{MeV})^2}\right)^{3/2}\nonumber\\
&&\nonumber\\
&\approx&(0.5-1.7)\times10^{-7}~~~~~\text{for}~~~~~\frac{\big|\theta_{a\eta_{_{ud}}}(\Delta u + \Delta d) + \sqrt{2}\,\theta_{a\eta_{_{s}}}\Delta s\big|}{10^{-4}}~\sim~(1.4-4).~~~~~~~~~~~~
\eeq

Here, despite uncertainties in (\ref{IsovectorRate8Be}), we are able to make the more firm prediction that, should an anomaly persist in the $^{\,8}\text{Be}^{*}(17.64)$ de-excitation rate at the level of $\Gamma_X/\Gamma_\gamma\approx \mathcal{O}(10^{-6})$, this would preclude our QCD axion variant from offering as a viable explanation of this excess.

\section{\label{Sec:AxionElectronCoupling} Axion-Electron Coupling}

Unlike its hadronic couplings, the axion coupling to electrons bears no consequence to the solution of the strong CP problem, and therefore it is a model dependent parameter that can be adjusted according to phenomenological constraints. Moreover, it is much less susceptible to calculational uncertainites, and can be probed in a variety of existing and upcoming experiments. In fact, it is conceivable that this MeV axion variant could be definitively excluded via its electron couplings before any substantial progress is made regarding its hadronic phenomenology. In this section, we will review existing bounds on the axion-electron coupling, and briefly discuss on-going experimental efforts to search for hidden photons that could test the MeV axion scenario as well.

\begin{figure}[t]
\begin{center}
\includegraphics[width=15cm]{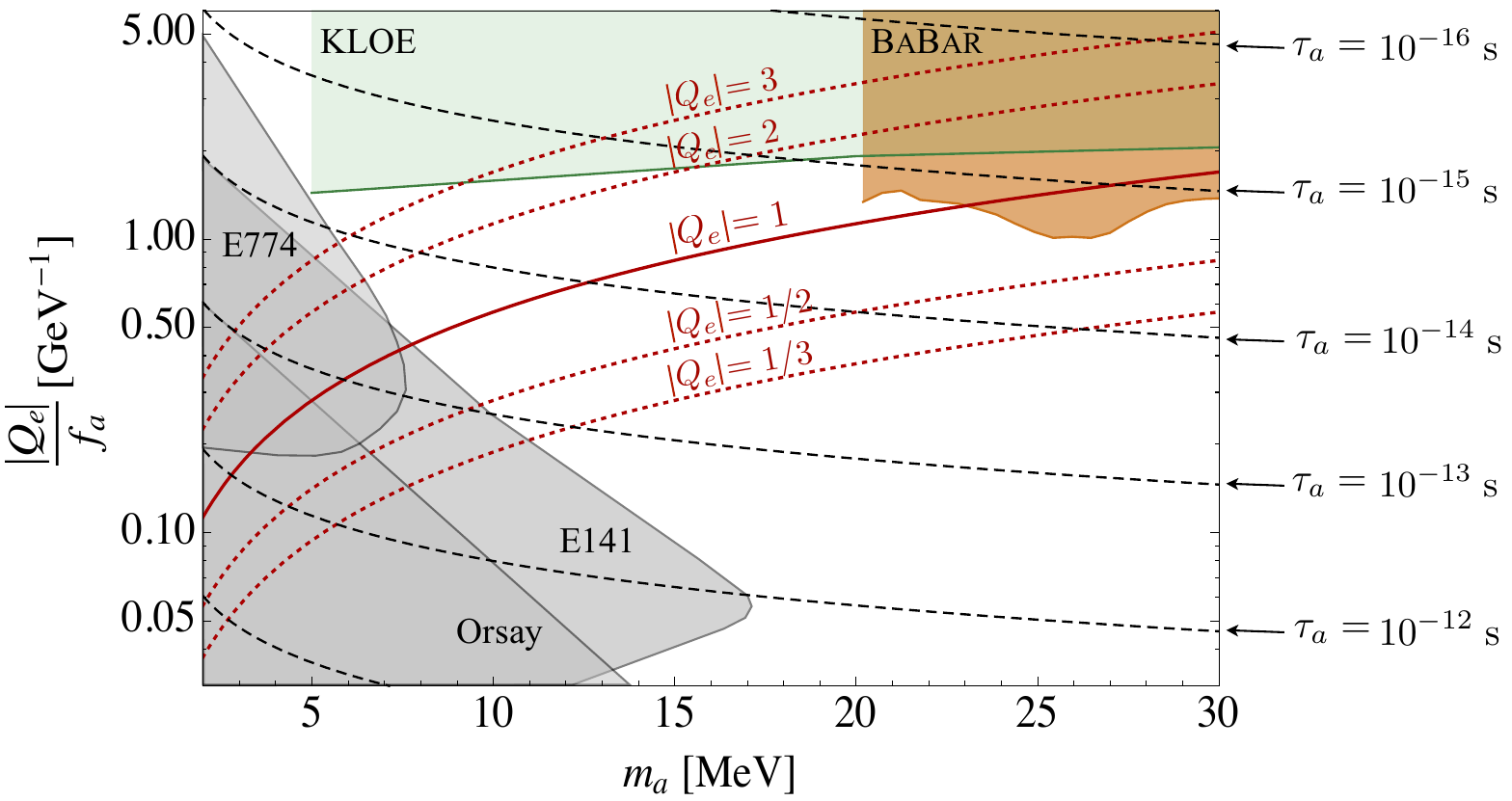}
\linespread{1}
\caption{\small Existing constraints on the axion-electron coupling for our specific axion variant. The red curves illustrate the $m_a$ vs. $|Q_e|/f_a$ relation for different values of $|Q_e|$. The black dashed lines are contours of the axion's lifetime. The shaded gray regions are excluded by beam dump experiments \cite{Bross:1989mp,Riordan:1987aw,Davier:1986qq}, and the shaded orange and green regions are excluded by BaBar \cite{Lees:2014xha} and KLOE \cite{Anastasi:2015qla} searches for dark photons, respectively. Constraints from the electron's anomalous magnetic moment are model dependent and therefore not shown in this plot (see Sec.\,\ref{eGminus2} for details).}
\label{MeVaxionPlot}
\end{center}
\end{figure}

For specificity, we reiterate the axion coupling to electrons:
\be
\mathcal{L}^\text{eff}_a~\supset~ \frac{Q_e}{f_a}\,m_e\;a\;\bar e\, i\gamma_5 e \,,
\ee
as well as the relation between the axion's mass and decay constant:
\beq\label{ma}
m_a~&=&~|Q_u+Q_d|\,\frac{\sqrt{m_um_d}}{~m_u+m_d}\,\frac{m_\pi f_\pi}{f_a}\nonumber\\
&\approx&~\frac{\sqrt{2}\;m_\pi f_\pi}{f_a}\;,
\eeq
and the axion lifetime, assuming that it is dominated by $a\rightarrow e^+e^-$:
\be\label{taua}
\tau_a^{-1}~&=&~\frac{m_a}{8\pi}\left(\frac{Q_e}{f_a}\,m_e\right)^2\,\sqrt{1-\frac{4\,m_e^2}{m_a^2}}\,.
\ee

The constraints discussed in this section are summarized in Fig.~\ref{MeVaxionPlot}.

\subsection{Beam dump constraints}
\label{BeamDumps}
In the 80's, several beam dump experiments have specifically targeted the QCD axion. Since the results of these searches require no re-interpretation, we refer the reader to the original papers for details on production and detection mechanisms. In this subsection, we compile the most significant contraints in the region of parameter space of interest.

Constraints from beam dumps can be avoided if axions are sufficiently short-lived so as to decay in the earth shielding, before reaching the detectors. Moreover, in order to remain experimentally viable, invisible decay modes of the axion must be subdominant by at least $\mathcal{O}(10^{-4})$ in order to avoid stringent constraints from $K^+\rightarrow\pi^+(a\rightarrow\text{invisible})$. In order to fulfill these requirements, the axion must be heavier than a few MeV and couple significantly to electrons.

 In Fig.~\ref{MeVaxionPlot}, we show contours of the axion lifetime as a function of the axion mass and coupling to electrons, $Q_e/f_a$. We also show the most relevant beam dump constraints as shaded gray regions, namely, FNAL E774 \cite{Bross:1989mp}, SLAC E141 \cite{Riordan:1987aw}, and Orsay \cite{Davier:1986qq}. Other overlapping but less powerful constraints, such as Bechis et al. \cite{Bechis:1979kp}, SLAC E56 \cite{Rothenberg:1972yr}, FNAL E605/772 \cite{Brown:1986xs,Guo:1989ia}, and KEK \cite{Konaka:1986cb} are omitted in order to make the plot more readable. Experiments with longer earth shieldings, such as SLAC E137 \cite{Bjorken:1988as} and CHARM \cite{Bergsma:1985qz} lie outside the plot.

\subsection{Constraints from $(g-2)_e$}
\label{eGminus2}
Next, we discuss the MeV axion contribution to the electron's anomalous magnetic moment. Constraints coming from $(g-2)_e$ are difficult to pin down  due to uncertainties in the axion's two-loop contribution, as well as the model dependence associated with UV completions of MeV axions. As we shall discuss in Sec.~\ref{GeVcompletion}, generic UV completions of such models contain additional particles that can also contribute substantially to $(g-2)_e$.

State-of-the-art calculations of $a_e\equiv(g-2)_e/2$ in the SM predict \cite{Aoyama:2014sxa}:
\be
a_e^\text{SM}~=~1\; 159\; 652\; 181.643\,(25)\,(23)\,(16)\,(763)\,\times\,10^{-12}\,, 
\ee
where the first three uncertainties are theoretical, and the last one stems from the error in the measurement of the fine structure constant.

The most precise measurementof $a_e$ to date \cite{Hanneke:2008tm,Hanneke:2010au}, on the other hand, gives:
\be
a_e^\text{exp}~=~1\; 159\; 652\; 180.73\,(0.28)\,\times\,10^{-12}\,,
\ee
and differs from the theoretical expectation in the SM by:
\be\label{eMDMrange}
a_e^\text{exp}-a_e^\text{SM}~=~-(0.91\pm 0.77\pm 0.28)\,\times\,10^{-12}\,.
\ee

The dominant contribution to this quantity from the MeV axion comes at one-loop and is given by:
\beq
\Delta a_e^\text{1-loop}&=&\;-\frac{2}{(4\pi)^2}\,\bigg(\frac{Q_e}{f_a}\,m_e\!\bigg)^2\,\bigg(\frac{m_e}{m_a}\bigg)^2\,\int_0^1\text{d}x~\frac{x^3}{(1-x)+\big(\frac{m_e}{m_a}\big)^2 x^2}\nonumber\\
&\approx&\;-\frac{Q_e^2}{(4\pi)^2}\;\bigg(\frac{m_e^2}{m_\pi\,f_\pi}\bigg)^2\,\left(\text{log}\bigg[\,\frac{m_a^2}{m_e^2}\,\bigg]-\frac{11}{6}\right)\,,
\eeq
where we have used (\ref{ma}) in the second equality.

The two-loop contribution (which is a Barr-Zee type diagram) suffers from large uncertainties due to the poor determination of $g_{a\gamma\gamma}$,
as well as the precise value of the cut-off scale~$\Lambda$ above which $g_{a\gamma\gamma}$ can no longer be treated as a point-like coupling \cite{Marciano:2016yhf}.
Setting $\Lambda\sim 4\pi f_\pi$, we have \cite{Melnikov:2006sr,Stockinger:2006zn,Giudice:2012ms}:
\beq
\Delta a_e^\text{2-loop}&\sim&\;\bigg(\frac{Q_e}{f_a}\,m_e\!\bigg)\;\frac{m_e}{\pi^2}\,\;g_{a\gamma\gamma}\;\, f_{PS}\bigg[\frac{4\pi f_\pi}{m_a}\bigg]\,,
\eeq
where, from (\ref{aFFdual}),
\be\label{gagg}
g_{a\gamma\gamma}~=~\frac{\alpha}{4\pi f_\pi}\bigg(\theta_{a\pi}+\frac{5}{3}\,\theta_{a\eta_{_{ud}}}+\frac{\sqrt{2}}{~3}\,\theta_{a\eta_{_{s}}}\!\bigg)\,,
\ee
and
\beq
f_{PS}[z]&=&\,\int_0^1\text{d}x~\frac{z^2/2}{x(1-x)-z^2}\,\log\bigg(\frac{x(1-x)}{z^2}\bigg)\\
&&\nonumber\\
&\approx&~\text{log}\,z + 1~~~~~\text{for}~~~~~z\gg1\,.\nonumber
\eeq

A few observations are in order: the one-loop contribution is negative and exceeds the allowed range in (\ref{eMDMrange}) by more than two standard deviations if $~|Q_e|\gsim1/2$. However, the two-loop contribution has the opposite sign if the product $Q_e\, g_{a\gamma\gamma}$ is positive, and could in principle partially cancel the one-loop contribution to a substantial degree. In particular, if the linear combination of hadronic mixing angles in (\ref{gagg}) is $\sim2\times10^{-2}$, then $Q_e\lsim1.3$ is still consistent with  (\ref{eMDMrange}) at 2$\sigma$. Note that this range of mixing angles, and therefore of $g_{a\gamma\gamma}$, is consistent with other experimental constraints, mainly because of the very poor determination of $\theta_{a\eta_{_{s}}}$. The rare decay $K^+\rightarrow \pi^+(a\rightarrow\gamma\gamma)$ is not yet sensitive to $\theta_{a\eta_{_{s}}}\lsim\mathcal{O}(10^{-1})$ (see Sec.~\ref{axion-photon}), and the contribution of $\theta_{a\eta_{_{s}}}$ to the $^{\,8}\text{Be}$ nuclear transitions is mild if $\Delta s \sim -0.02$, which is suggested by recent lattice results \cite{QCDSF:2011aa, Engelhardt:2012gd, Abdel-Rehim:2013wlz, Bhattacharya:2015gma, Abdel-Rehim:2015owa, Abdel-Rehim:2015lha, Green:2017keo}.

Given all these considerations, existing measurements of the electron anomalous magnetic moment cannot provide a robust and model independent exclusion of the MeV axion parameter space, and therefore we do not include it in Fig.~\ref{MeVaxionPlot}.

\subsection{Constraints from searches for dark photons}
\label{darkphoton}

Over the past decade, there has been a spur in phenomelogical studies of light, weakly coupled vectors (``dark photons''), mostly motivated by dark matter phenomenology \cite{Essig:2009nc,Bjorken:2009mm,Batell:2009yf,Batell:2009di}. As a result, many experimental proposals have been put forth to search for dark photons in the near future \cite{Hewett:2012ns,Essig:2013lka,Alekhin:2015byh,Alexander:2016aln}. As it turns out, dark photon production and detection mechanisms are very similar to those of light pseudoscalars such as the axion, and several constraints on dark photons decaying visibly can be easily recast as limits on the MeV axion.

In this section, we briefly comment on two experimental strategies that are relevant for MeV axions, namely, production in $e^+e^-$ collisions, and at fixed-target experiments. We also translate existing bounds as well as future projections of proposed experiments in the MeV axion parameter space.

The axion can be produced in association with a photon in $e^+e^-$ annihilation via the $a\,\bar e i\gamma_5 e$ vertex. In the limit that the center-of-mass energy $\sqrt{s}\,\gg m_a$, the cross-section computed at leading order is:
\beq\label{babarXsec}
\frac{\;\text{d}\sigma(e^+e^-\rightarrow \gamma\, a)\;}{\text{d}\cos\theta_\gamma}~=~\bigg(\frac{Q_e}{f_a}\,m_e\bigg)^2\,\frac{\alpha}{~2s\,\sin^2\theta_\gamma+\frac{m_e^2}{2}~}\,,
\eeq
where $\theta_\gamma$ is the angle of the final state photon with respect to the beam axis, and we show the explicit dependence on $m_e$ that regulates the collinear divergence when $|\text{sin}\theta_\gamma|\rightarrow 0$.

Recent results from BaBar \cite{Lees:2014xha} and KLOE \cite{Anastasi:2015qla} looking for $e^+e^-\rightarrow \gamma\, (A^\prime\rightarrow e^+e^-)$ can be easily recast in the axion parameter space using (\ref{babarXsec}). We display the corresponding limits in Fig.~\ref{MeVaxionPlot}.

The axion can also be produced in fixed target experiments via ``axion-bremsstrahlung'' when an energetic electron scatters off of a heavy nuclear target. The differential cross-section as a function of the axion emission angle with respect to the beam, $\theta_a$, and the fraction $x_a$ of the incoming electron energy carried by the radiated axion, $x_a\equiv E_a/E_0$, has been worked out in \cite{Tsai:1986tx}. Below, we just quote the result:
\beq\label{FTdiffXcos}
\frac{\;\text{d}\sigma(eN\rightarrow eN a)\;}{\text{d}x_a\;\text{d}\cos\theta_a}\,=\frac{\alpha^2}{\pi}\,\bigg(\frac{Q_e}{f_a}\,m_e\bigg)^2 \beta_a\,E_0^2\,\frac{x_a}{U^2}\,\bigg(\frac{x_a^2}{2}-\frac{m_a^2}{U^2}\,x_a(1-x_a)(E_0^2\,\theta_a^2\,x_a)  \bigg)\,\chi\,,~~~
\eeq
where
\beq
U(x_a,\theta_a)~\equiv~E_0^2\,\theta_a^2\,x_a+m_a^2\,\frac{1-x_a}{x_a}+m_e^2x_a~~~,~~~ \beta_a\equiv\sqrt{1-m_a^2/E_0^2}\,,
\eeq
and $\chi$ encodes nuclear form factors (in the notation of \cite{Bjorken:2009mm}, $\chi=Z^2\,\mathcal{L}og$). For further details, see \cite{Tsai:1973py,Kim:1973he,Tsai:1986tx}.

Much like the case of dark photon bremsstrahlung, axion emission is almost collinear, and its characteristic emission angle is parametrically smaller than the angle of its decay products with respect to the incident beam \cite{Bjorken:2009mm}. Integrating (\ref{FTdiffXcos}) over emission angles, we obtain
\beq\label{FTdiffX}
\frac{\;\text{d}\sigma(eN\rightarrow eN a)\;}{\text{d}x_a}~&=&~\frac{\alpha^2}{3\pi\,m_a^2}\,\bigg(\frac{Q_e}{f_a}\,m_e\bigg)^2 \beta_a\,\chi\,\frac{x_a^3}{~(1-x_a)+\big(\frac{m_e}{m_a}\big)^2\,x_a^2~}\,,
\eeq
which is more peaked towards $x_a=1$ than the corresponding $x_{A^\prime}$ dependence of dark photon emission.

\begin{figure}[t]
\begin{center}
\includegraphics[width=15cm]{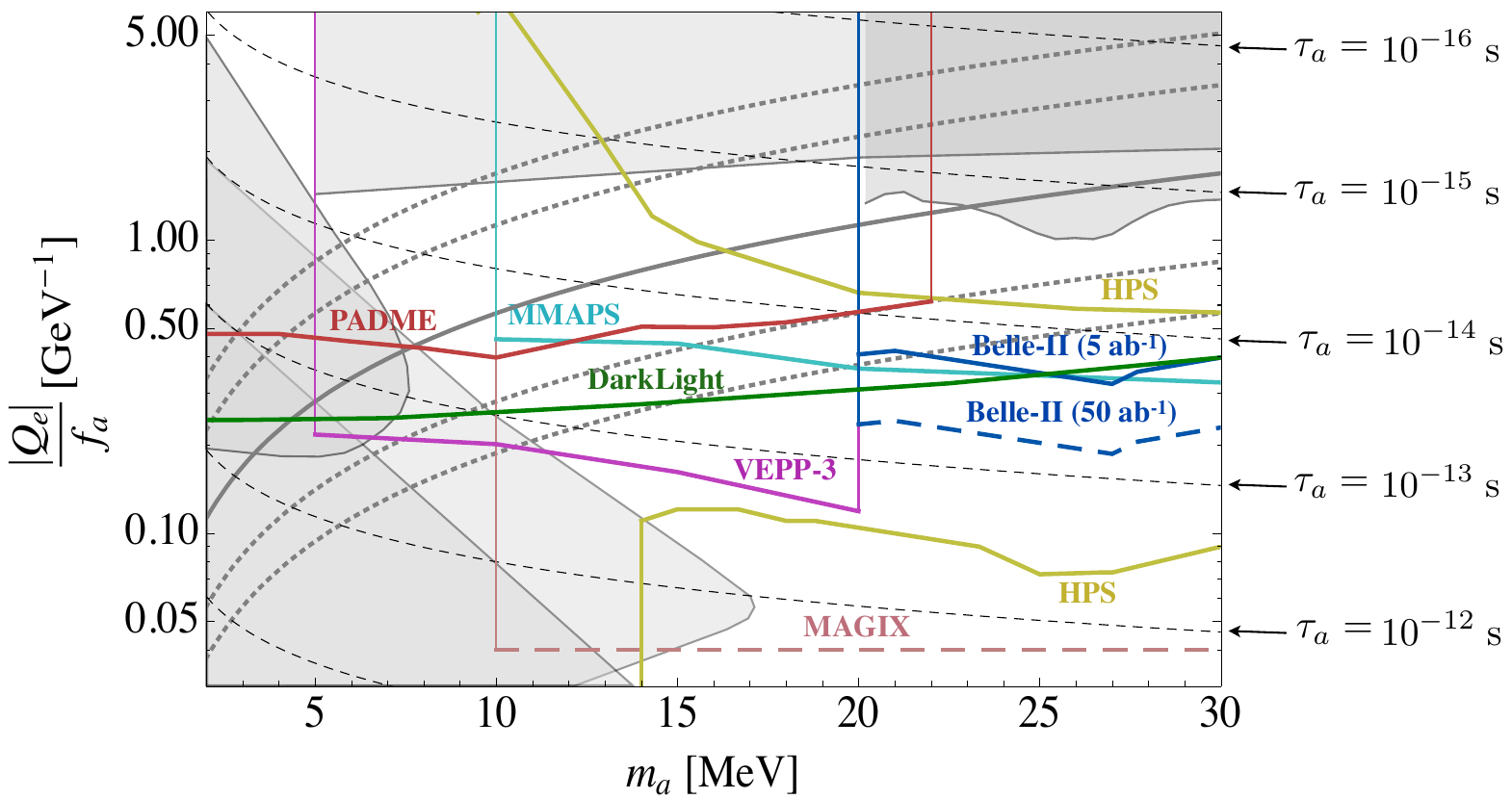}
\linespread{1}
\caption{\small Projected reach of several planned dark photon experiments into the MeV axion parameter space, based on \cite{Alexander:2016aln}. Gray curves and gray shaded areas are from Fig.~\ref{MeVaxionPlot}.}
\label{eeFuture}
\end{center}
\end{figure}

Finally, we quote the total production cross-section integrated over angle and energy:
\beq\label{FTtotal}
\sigma(eN\rightarrow eN a)\,&=&\;\frac{\alpha^2}{3\pi\,m_a^2}\,\bigg(\frac{Q_e}{f_a}\,m_e\bigg)^2 \chi\;\left(\log\bigg[\frac{1}{(1-x_a)_c}\bigg]\,-\,\frac{11}{6}\,+\,\mathcal{O}\bigg(\frac{m_a^2}{E_0^2}\bigg)\right)~~~~~~\nonumber\\
&\approx&\,\frac{\alpha^2\,Q_e^2}{6\pi}\;\bigg(\frac{m_e}{\,m_\pi\,f_\pi\,}\bigg)^2 \chi\,\left(\log\bigg[\frac{1}{(1-x_a)_c}\bigg]\,-\,\frac{11}{6}\,+\,\mathcal{O}\bigg(\frac{m_a^2}{E_0^2}\bigg)\right),~~~~~~
\eeq
where
\beq
(1-x_a)_c~\equiv~\max\bigg(\frac{m_e^2}{m_a^2}\,,\,\frac{m_a^2}{E_0^2}\bigg)\,,
\eeq
and we have used (\ref{ma}) in the second equality. We see, therefore, that the total cross-section for axion emission increases logarithmically with $m_a$ while $m_a\lesssim\sqrt{m_e\,E_0}$, whereas it decreases logarithmically with $m_a$ in the regime $\sqrt{m_e\,E_0}\lesssim m_a\ll E_0$.

In Fig.~\ref{eeFuture}, we translate the projected reach of several planned dark photon experiments into the MeV axion parameter space, using the following rule of thumb relating the dark photon kinetic mixing parameter $\epsilon_{A^\prime}$ and the axion-electron coupling that would yield comparable signal strengths:
\be\label{epsilonQemap}
~~~\epsilon_{A^\prime}^2\,\alpha~~~\sim~~~\frac{1}{4\pi}\,\bigg(\frac{Q_e}{f_a}\,m_e\bigg)^2\,.
\ee
We emphasize that (\ref{epsilonQemap}) is only approximate, since the angular dependence of pseudoscalar cross-sections differs from that of vectors. Fig.~\ref{eeFuture} includes ongoing and proposed dark photon searches via $e^+e^-$ annihilation (VEPP-3 at BINP \cite{Wojtsekhowski:2009vz,Wojtsekhowski:2012zq}, PADME in Frascati \cite{Raggi:2014zpa}, MMAPS at Cornell \cite{Alexander:2016aln} and Belle-II at KEK \cite{TheBelle:2015mwa}), as well as dark photon bremsstrahlung from electrons scattering off of heavy fixed targets (HPS \cite{Battaglieri:2014hga} and DarkLight \cite{Balewski:2013oza,Balewski:2014pxa} at JLab, MAGIX \cite{Denig:2016dqo} at Mainz). These projections were based on the {\it Dark Sectors 2016 Workshop} community report \cite{Alexander:2016aln}. We refer the reader to this report for further details.

\section{Other Constraints}
\label{other}

Finally, we comment on other potentially constraining observables that could probe the MeV axion parameter space. We discuss them in lesser detail because the resulting limits are either weaker than the ones previously discussed, suffer from similar hadronic uncertainties, or lack experimental information specific to the axion related signature.

Bounds from the hyperfine splitting between the $1 ^{3\!}S_1$ and $1^{1\!}S_0$ positronium levels \cite{Schaefer:1989tv,Ishida:2013waa}, for instance, are currently not competitive with beam dumps contraints shown in Fig.~\ref{MeVaxionPlot}. Neither are bounds from neutron-nucleus and neutron-electron scattering \cite{Barbieri:1975xy, Nesvizhevsky:2007by}, since the axion mediated contribution is spin-dependent \cite{Kamyshkov:2008qq,Daido:2017hsl}.

Competitive constraints on axion hadronic mixing angles could be potentially obtained from dedicated analyses of existing data on other rare meson decays, such as $K^+\rightarrow \ell^+\nu_\ell\, a\,$, $K^0_L\rightarrow \pi\pi a\,$, $K^0_S\rightarrow \pi^0 a\,$, and $\eta,\eta^\prime\rightarrow\pi\pi a\,$. The relevant final states for these observables have been measured in high statistics samples \cite{Poblaguev:2002ug,Lai:2003ad,Adams:1998eu,AlaviHarati:2002nf,Batley:2003mu,Lai:2001jf,Barr:1993te,Bargholtz:2006gz,Akhmetshin:2000bw,Berlowski:2007aa,Ambrosino:2008cp,Naik:2008aa,Ablikim:2013wfg}, but, to the best of our knowledge, no dedicated searches for $e^+e^-$ resonances in the few MeV invariant mass range have been performed.

Other less sensitive rare decays also exist, which could become competitive if the branching ratio sensitivity were substantially improved. Examples are $B$ meson decays\footnote{Note that EW penguins with charm or top quarks in the loop do not contribute in this case, because the axion does not couple to heavy quarks.},  $B\rightarrow K a$ \cite{Aubert:2008ps,Wei:2009zv}; radiative quarkonia decays, $J/\psi,\Upsilon,\Phi\rightarrow\gamma a\,$ \cite{Bowcock:1986ig,Mageras:1986nz,Albrecht:1986ht,Armstrong:1996hg,Druzhinin:1987nx,Dimova:2008zz}; hyperon decays, such as $\Sigma^+\rightarrow p^+ a\,$ \cite{Suzuki:1986dx}; and the not-yet-observed decays $\eta,\eta^\prime\rightarrow e^+e^-$ \cite{Agakishiev:2013fwl,Achasov:2015mek}, which are highly suppressed in the SM.

We emphasize that axion production via its coupling to photons\footnote{See \cite{Dolan:2017osp} for a recent study of ALP photo-production.} should be subdominant to the processes discussed in Sec.~\ref{darkphoton} induced by the axion's electronic couplings. This can be easily seen by noting that the effective axion-photon coupling, $g_{a\gamma\gamma}\,a\,F\tilde{F}$, is suppressed by an effective scale of $\mathcal{O}(10-100)\,\TeV$:
\be
g_{a\gamma\gamma}~=~\frac{\alpha}{\,4\pi f_\pi}\,\left(\theta_{a\pi}+\frac{\theta_{a\eta_8}}{\sqrt{3}}+2\,\frac{\theta_{a\eta_0}}{\sqrt{3/2}}\right)~\sim~\frac{\,\mathcal{O}(0.1-1)\,}{\,10\,\TeV\,}\,.
\ee
A comparable effective scale suppresses the couplings of the axion to $Z\gamma$, thereby suppressing the rate for the decay $Z\rightarrow\gamma a$ below current experimental bounds. This rare $Z$ decay is typically a sensitive probe of generic ``axion-like particles'' (ALPs) with low decay constants (see recent studies in \cite{Dolan:2017osp,Dror:2017ehi,Dror:2017nsg}). A generic ALP $\phi$ with decay constant $f_\phi$ couples to $\gamma\,Z$ with typical strength $g_{\phi \gamma Z}\sim\frac{\alpha}{4\pi f_\phi}$. This parametric dependence does not apply for an axion with $f_a\sim\mathcal{O}(\GeV)$ for the following reason: in the case of a generic ALP, the $\phi$-$\gamma$-$Z$ coupling is generated by integrating out electroweakly charged fermions heavier than $M_Z$. An MeV axion, on the other hand, is typically precluded from coupling to such heavy fermions by the fact that the PQ symmetry must be preserved down to the scale of $f_a\sim\mathcal{O}(1-10)\;\GeV$. In other words, the MeV axion can only have perturbative couplings to fermions with mass $m_f\;=\;y_f\,\mathcal{O}(f_a)\;\lesssim\; \mathcal{O}(f_a)\;\ll\; M_Z$. Loops of such light fermions induce much smaller $g_{a\gamma Z}$ couplings, $g_{a\gamma Z}\;\sim\; \frac{m_f}{M_Z}\,\frac{\alpha}{4\pi f_a}$.

So far we have assumed that the MeV axion does not couple to neutrinos. This can be naturally realized if neither the lepton doublets nor right-handed Dirac and/or sterile neutrinos carry PQ charge. If we relax this assumption, the typical axion coupling to neutrinos would scale as $m_\nu/f_a\sim\mathcal{O}(10^{-11}-10^{-10})$, which is likely too small to be phenomenologically interesting. On the other hand, PQ-charged {\it sterile} neutrinos could couple to the axion much more strongly, namely, with strength $\mathcal{O}(m_{\nu_s}/f_a)$. If $m_{\nu_s}$ is heavy enough, this could lead to observable signatures. A particularly well motivated mass range for sterile neutrinos is $m_{\nu_s}\sim\mathcal{O}(1-10)$ keV, where sterile neutrinos can constitute warm dark matter and potentially address structure formation problems such as the core vs cusp and missing satelites problems (see \cite{Adhikari:2016bei} and references therein). In this mass range, the branching ratio for axion decay into sterile neutrinos would be:
\beq
\label{atonunu}
\Br(a\rightarrow \nu_s \nu_s)~\sim~\mathcal{O}\bigg(\frac{m_{\nu_s}^2}{m_e^2}\bigg)~\sim~\mathcal{O}(10^{-4} - 10^{-6}).
\eeq 
An interesting consequence of (\ref{atonunu}) would be a contribution to the rare kaon decay $K^+\rightarrow\pi^+a\rightarrow\pi^+\nu_s\nu_s$ at the level of:
\beq
\Br(K^+\rightarrow\pi^+\nu_s\nu_s)~\sim~\mathcal{O}(10^{-10} - 10^{-12})\;\left(\frac{\Br(K^+\rightarrow\pi^+a)}{10^{-6}}\right)\,,
\eeq
which is compatible with the strongest bounds set by BNL's E787/E949 experiments \cite{Adler:2002hy,Adler:2004hp,Anisimovsky:2004hr}, namely,  $\Br(K^+\rightarrow\pi^+(X^0\rightarrow\text{invisible}))\lesssim 0.45\times 10^{-10}$ for $m_{X^0}\lesssim 70$ MeV.  CERN's NA62 experiment will soon supersede these bounds by at least one order of magnitude, since it is expected to have sensitivity to the SM prediction of $\Br(K^+\rightarrow\pi^+\nu\bar\nu)|_{_\text{SM}}=(8.4\pm1.0)\times 10^{-11}$ \cite{Buras:2015qea} with better than 10\% accuracy \cite{Corvino:2018sjc}.

Other interesting possibilities for neutrino phenomenology could come from off-diagonal couplings of the axion to active ($\nu_\ell$) and sterile neutrinos ($\nu_s$), such as $\lambda\,a\,\nu_{\ell\,}\nu_s $, especially if the sterile neutrinos are light enough to be produced in laboratory experiments and/or astrophysical processes. Effects such as $\nu_s\rightarrow \nu_\ell\, a\rightarrow \nu_\ell \,e^+e^- $ (if $m_{\nu_s}>m_a$), non-standard neutrino interactions, MSW-type resonance effects in neutrino propagation in matter, and neutrino transport in core-collapse supernova are interesting directions to explore, but are beyond the scope of this paper.

\section{\label{GeVcompletion} GeV Scale Completions of MeV Axion Models}

The Peccei-Quinn breaking scale $f_a$ suggests that there is new dynamics around $1-10$ GeV. While there are many constraints on what type of new particles may be associated with this new dynamics, there is still a large degree of model dependence in UV completions of the MeV axion. A thorough exploration of the viable phenomelogy is beyond the scope of this paper, and will be deferred to a future publication \cite{future}. In this section, we briefly illustrate a few possibilities with a simple toy-model.

Consider introducing two new complex scalar degrees of freedom, $\Phi_u$ and $\Phi_d$, with PQ charges $\Qpq_{\Phi_u}=Q_u=2$ and $\Qpq_{\Phi_d}=Q_d=1$, respectively. We can then UV complete the PQ mechanism at the GeV scale by writing:
\be\label{PQGeV}
\mathcal{L}_{\text{PQ}}^{\Lambda=\text{GeV}}~~\supset~-\big(\,y_u\,\Phi_u\,uu^c~+~y_d\,\Phi_d\,dd^c~+~\text{h.c.}\,\big) ~-~ V(\Phi_u,\Phi_d)\,.
\ee
The scalar potential $V(\Phi_u,\Phi_d)$ enforces the PQ symmetry, and induces vacuum expectation values for $\Phi_u$ and $\Phi_d$, hence breaking the PQ symmetry:
\beq\label{PhiVEVs}
\langle\Phi_u\rangle&=&\frac{f_u}{\sqrt{2}}~=~\frac{m_u}{y_u}\,,\\
\langle\Phi_d\rangle&=&\frac{f_d}{\sqrt{2}}~=~\frac{m_d}{y_d}\,.
\eeq

We can then decompose $\Phi_u$ and $\Phi_d$ into real scalar and pseudoscalar components:
\beq
\Phi_u &=& \left(\frac{f_u}{\sqrt{2}}\,+\,\frac{\varphi_u}{\sqrt{2}}\right) \,\text{Exp}\; i\bigg(Q_u\,\frac{a}{f_a} \,+\, \frac{Q_d}{\tan\beta_{_\text{PQ}}}\frac{\eta_{_\text{PQ}}}{f_a}\bigg)\;,\\
\Phi_d &=& \left(\frac{f_d}{\sqrt{2}}\,+\,\frac{\varphi_d}{\sqrt{2}}\right) \,\text{Exp}\; i\bigg(Q_d\,\frac{a}{f_a}\,-\,Q_u\tan\beta_{_\text{PQ}}\frac{\eta_{_\text{PQ}}}{f_a}\bigg)\;.
\eeq
Above, $\tan\beta_{_\text{PQ}}\equiv f_u/f_d\,$, the pseudoscalar $a$ is the MeV axion, and the axion decay constant, $f_a$, is given by $f_a^2~\equiv~Q_u^2\, f_u^2 + Q_d^ 2\,f_d^2 ~=~ 4f_u^2+f_d^2 $.

In this toy model, we have introduced 3 extra degrees of freedom besides the MeV axion: two real scalars, $\varphi_u$ and $\varphi_d$, and one pseudoscalar, $\eta_{_\text{PQ}}$. The natural scale for their masses is set by $f_a$. Such light states must, therefore, be electroweak singlets in order to be phenomenologically viable. As a consequence, the couplings of $\Phi_u$ and $\Phi_d$ to SM fermions in (\ref{PQGeV}) descend from higher dimensional operators and must be generated after electroweak symmetry breaking. We shall return to this point in the next section.

Because of their couplings to light quarks, the new states $\varphi_u$, $\varphi_d$, and $\eta_{_\text{PQ}}$ are produced in hadronic interactions, and decay dominantly to hadrons -
though with smaller cross-sections and narrower widths than typical QCD hadronic resonances.
Unfortunately, estimating their widths and mixings with QCD resonances is challeging, since they lie in a regime where neither perturbative QCD nor chiral perturbation methods are reliable. It is conceivable, at least in principle, that these states are not excluded. The scalar resonances $\varphi_u$ and $\varphi_d$, for instance, might not have been identified if lying in the murky mass range below 2~GeV \cite{Amsler}, where many broad scalar resonances overlap and might constitute a large and complicated background to disentangle. The same argument is less likely to apply for the pseudoscalar resonance $\eta_{_\text{PQ}}$, unless it lies in a mass range where the hadronic pseudoscalar spectrum is poorly understood. For instance, the 1300-1500 MeV range contains three $0^{-+}$ states, namely, $\eta(1295)$, $\eta(1405)$ and $\eta(1475)$. The exact nature of these states is still subject to debate \cite{Amsler2}, with interpretations ranging from those being radial excitations of lighter pseudoscalars \cite{Close:1997pj,Barnes:1996ff,Gutsche:2008qq} 
 to pseudoscalar glueballs \cite{Close:1996yc,Farrar:1996ye,Li:2003cn,Faddeev:2003aw,Masoni:2006rz,Li:2007ky,Cheng:2008ss,Gutsche:2009jh,Li:2009rk}. Some authors dispute the existence of the $\eta(1295)$ state and claim that there is a single pseudoscalar meson in this mass range, the $\eta(1440)$ state, which would be the first radial excitation of the $\eta$ \cite{Klempt:2007cp,Klempt:2006jk}. There are also claims that the $\eta(1405)$ and $\eta(1475)$ structures might originate from a single pole \cite{Aceti:2012dj}, the splitting being due to nodes in the decay amplitudes \cite{Klempt:2007cp}, or amplitude mixing via a triangular singularity \cite{Wu:2011yx,Wu:2012pg}. Whatever the nature of $\eta(1295)$, $\eta(1405)$ and $\eta(1475)$ may be, a priori they might present a challenging background for $\eta_{_\text{PQ}}$, and possibly mix substantially with it. 
A less speculative and more careful investigation of these possibilities will be done in \cite{future}.

We conclude by commenting on the coupling of these states to electrons. One possibility is that either $\Phi_u$ or $\Phi_d$ have electron Yukawa couplings, in which case all real degrees of freedom (namely, $\varphi_u$, $\varphi_d$, $\eta_{_\text{PQ}}$ and $a$) will couple to electrons (either directly or through mixing), with typical strength $\mathcal{O}(m_e/f_a)$. Alternatively, we may introduce a third PQ-charged complex scalar, $\Phi_e=\varphi_e\, e^{i\,\eta_e/f_e}$, which will be responsible for generating the electron mass and the axion-electron coupling. Mixings between the $\Phi_e$ and $\Phi_{u,d}$ degrees of freedom will likewise result in couplings of $\varphi_{u,d,e}$ and $\eta_{_\text{PQ},e}$ to electrons, although in this case some of the couplings might be suppressed or enhanced relative to $m_e/f_a$. Depending on the details of a specific model, these states might give important contributions to $(g-2)_e$, $\Gamma(\pi^0\rightarrow e^+e^-)$, and might be searched for in $e^+ e^-$ annihilation and fixed target experiments as well. A phenomenological study will be considered in \cite{future}.

\section{\label{EWcompletion} EW Scale Completions of MeV Axion Models}

As discussed in the previous section, the dynamics that breaks the PQ symmetry at the GeV scale generically requires new light degrees of freedom coupling to $\SU(2)_W$-charged quark bilinears. However, these new light particles themselves must not be $\SU(2)_W$-charged, otherwise they would be excluded, for instance, by measurements of the $Z^0$ width. Therefore, the Yukawa couplings of the PQ sector to SM fermions violate $\SU(2)_W\times\U(1)_Y$ and can only arise from higher dimensional operators after EW symmetry breaking. These operators can be generated by integrating out new degrees of freedom at the EW scale, such as heavy PQ-charged scalar doublets, or heavy vector-like quarks, leading to interesting and distinct LHC signatures. Although a thorough exploration of all possibilities is beyond the scope of this paper, we will brielfy consider a simple EW completion of the toy-model discussed in Sec.\ref{GeVcompletion}, and comment on the associated phenomenology.

For each EW singlet PQ scalar $\Phi_f$, we introduce a new $\SU(2)_W$ doublet $H_f$ with the same corresponding PQ charge, $\Qpq_{H_f}=\Qpq_{\Phi_f}$. We can then write EW preserving Yukawa couplings, $H_f f_L f^c$, and tri-scalar ``$A$-terms'':
\be\label{PQEW}
\mathcal{L}_{\text{PQ}}^{\Lambda=\text{EW}}~~\supset~&-&\big(\,Y_u\,H_u\,Q_1 u^c~+~Y_d\,H_d\,Q_1 d^c~+~Y_e\,H_e\,L_1 e^c~+~\text{h.c.}\,\big)\nonumber\\
&-&\big(\,A_u\, \Phi_u^\dagger H_u H_{_\text{SM}}^\dagger~+~A_d\, \Phi_d^\dagger H_d H_{_\text{SM}}~+~A_e\, \Phi_e^\dagger H_e H_{_\text{SM}}   ~+~\text{h.c.}\,\big)\,.
\ee
Above, the $A_f$ coefficients have dimensions of mass, $H_{_\text{SM}}$ is the doublet responsible for EW symmetry breaking and contains the 125 GeV Higgs, and $H_e$, $\Phi_e$ may be independent fields, or may be identified with $H_d$, $\Phi_d$ or $H_u^\dagger$, $\Phi_u^\dagger$.

Giving large masses $m_{H_f}\gsim\mathcal{O}(100~\GeV)$ for the new PQ doublets $H_f$, we can integrate them out and obtain the effective interactions below the EW symmetry breaking scale:
\be
\mathcal{L}_{\text{PQ}}^{\Lambda=\GeV}~~\supset~&-&\big(\,y_u\,\Phi_u\,uu^c~+~y_d\,\Phi_d\,dd^c~+~y_e\,\Phi_e\,e e^c~+~\text{h.c.}\,\big)
\ee
where
\be\label{YukawaRelations}
y_f~\equiv~Y_f\;\frac{\,A_f\,\langle H_{_\text{SM}}\rangle\,}{m_{H_f}^2}\,,
\ee
with $\langle H_{_\text{SM}}\rangle=174~\GeV$.

A parameter of particular relevance for phenomenology is the mixing angle between the light singlet $\Phi_f$ and the neutral component of the heavy doublet $H_f^0$:
\be
\theta^f_{\Phi H}~\equiv~\frac{\,A_f\,\langle H_{_\text{SM}}\rangle\,}{m_{H_f}^2}\,.
\ee
Firstly, this parameter quantifies the degree of tree-level tuning required to maintain the hierarchy between the singlet states (at the GeV scale) and the doublet states (at the EW scale). A simple measure of this fine-tuning is given by:
\be\label{PhiHtuning}
\text{F.T.}~\sim~\frac{1}{\theta^{f\;2}_{\Phi H}}\,\frac{m_{\Phi_f}^2}{m_{H_f}^2}~\sim~\frac{\mathcal{O}(10^{-4})}{\theta^{f\;2}_{\Phi H}}\,.
\ee
Imposing a tolerance of at most 10\% tuning, (\ref{PhiHtuning}) requires $\theta^f_{\Phi H}\lsim 0.03$.

Secondly, $\theta^f_{\Phi H}$ indirectly determines the coupling of $H_f$ to quarks. Considering the up quark for concreteness, we have from (\ref{YukawaRelations}) and (\ref{PhiVEVs}),
\be
Y_u~=~\sqrt{2}\;\frac{m_u}{f_u}\,\frac{1}{\theta^u_{\Phi H}}~\approx~ 0.1\times\bigg(\frac{0.03}{\theta^u_{\Phi H}}\bigg)\,\bigg(\frac{\GeV}{f_u}\bigg)\,.
\ee
Such sizable Yukawa coupling would lead to a large production of $H_u$ at the LHC, with 13 TeV cross-sections ranging from $\mathcal{O}(1-10^3)$~pb~~for~~ $\theta^f_{\Phi H}\sim(0.01-0.03)$~~and~~$m_{H_u}\sim(100-500)$~GeV. Existing LHC searches for leptophobic vectors ($Z^\prime$) \cite{Dobrescu:2013coa} can already place non-trivial upper bounds on processes such as $pp\rightarrow H_u(\rightarrow jj)+j$ \cite{Sirunyan:2017nvi,Khachatryan:2016ecr,Aaboud:2018zba} and $pp\rightarrow H_u(\rightarrow jj)+\gamma$ \cite{Aaboud:2018zba,ATLAS:2016bvn}, ranging from $Y_u\lesssim (0.1-0.4)$ depending on $m_{H_u}$. Similar considerations hold for $H_d$ production.


\begin{table}
\begin{center}
\resizebox{1 \textwidth}{!}{ 
    \begin{tabular}{|c|c|c|}
	\hline
	 Rare Decay  & Potential Signature & Signal Strength   \\
	\botrule
	 ~$pp~\rightarrow~ H_{u,d}^0~\rightarrow~ Z^0a$ ~&~~ $Z+\gamma^*~, ~Z+LJ$ ~~&~ $\sigma\sim\mathcal{O}(10^{-3}-1)$~pb ~\\
	\hline
	~$pp~\rightarrow~ H_{u,d}^\pm~\rightarrow~ W^\pm a$ ~&~~ $W+\gamma^*~,~W+LJ$ ~~&~ $\sigma\sim\mathcal{O}(10^{-3}-1)$~pb ~\\
	\hline
	 ~$pp~\rightarrow~ H_{u,d}^0~\rightarrow~ h\;\varphi_{u,d}/\eta_{_\text{PQ}}$ ~&~~ $h+j~, ~h+P\!J~,~h+\tau_h$ ~~&~ $\sigma\sim\mathcal{O}(0.1-1)$~pb ~\\
	\hline
	~$pp~\rightarrow~ H_{u,d}^0~\rightarrow~ Z^0\,\varphi_{u,d}/\eta_{_\text{PQ}}$ ~&~~ $Z+P\!J~,~Z+\tau_h$ ~~&~ $\sigma\sim\mathcal{O}(10^{-3}-1)$~pb ~\\
	\hline
	~$pp~\rightarrow~ H_{u,d}^\pm~\rightarrow~ W^\pm\,\varphi_{u,d}/\eta_{_\text{PQ}}$ ~&~~ $W+P\!J~,~W+\tau_h$ ~~&~ $\sigma\sim\mathcal{O}(10^{-3}-1)$~pb ~\\
	\hline
	 ~ $Z^0\rightarrow a\,\phi_{u,d}$ ~&~~ $\Gamma_Z~,~Z\rightarrow \gamma^*\,j~,~Z\rightarrow \gamma^*\, P\!J~,~Z\rightarrow LJ\,P\!J$  ~~&~  $\text{Br}\sim \mathcal{O}(10^{-10}-10^{-8})$  ~\\
	\hline
	 ~ $Z^0\rightarrow \eta_{_\text{PQ}}\,\phi_{u,d}$ ~&~~ $R_Z~,~Z\rightarrow j\, P\!J~,~Z\rightarrow P\!J\, P\!J$  ~~&~  $\text{Br}\sim \mathcal{O}(10^{-10}-10^{-8})$  ~\\
	\hline
	~ $h\rightarrow \phi_{u,d}\,\phi_{u,d}\,,~\eta_{_\text{PQ}}\,\eta_{_\text{PQ}}$ ~&~~ $h\rightarrow j\,P\!J~,~h\rightarrow P\!J\, P\!J~,~h\rightarrow \tau_h \tau_h$  ~~&~  $\text{Br}\sim \mathcal{O}(10^{-6}-10^{-3})$  ~\\
	\hline
	~ $h\rightarrow H^0_{u,d}\;\varphi_{u,d}/\eta_{_\text{PQ}}~(m_{H_{u,d}}<m_h)$ ~&~~ $h\rightarrow (jj)\,j~,~h\rightarrow (jj)\, P\!J~,~h\rightarrow (jj)\, \tau_h$  ~~&~  $\text{Br}\sim \mathcal{O}(10^{-3}-10^{-2})$  ~\\
	\hline
    \end{tabular}
    }
    \end{center}
    \caption{\label{RareDecayTable} Compilation of rare decays that could potentially probe the model in (\ref{PQEW}). The middle column shows potential signatures, {\it assuming} that the MeV axion $a$ would be tagged either as a converted photon ($\gamma^*$) or a lepton-jet ($LJ$); and that $\varphi_{u,d}\,,\;\eta_{_\text{PQ}}$ would be tagged either as a jet ($j$), a photon-jet ($PJ$), or a hadronic tau ($\tau_h$). The third column shows typical ranges for signal cross-sections/branching ratios assuming $\theta^f_{\Phi H}\sim(0.01-0.03)$~~and~~$m_{H_u}\sim(100-300)$~GeV.}
\end{table}

Finally, $\theta^f_{\Phi H}$ controls a variety of rare decays of the 125 GeV Higgs h, $Z^0$ and $H_f$ to final states with PQ scalars $a,\,\eta_{_\text{PQ}},\,\varphi_{u},\,\varphi_{d}$. We compile some of these possibilities in Table~\ref{RareDecayTable}. The reach of existing measurements and future search strategies for these rare decays will depend on how the boosted PQ scalars will be tagged by LHC detection algorithms. Typically, the GeV states $\varphi_{u},\,\varphi_{d}$, $\eta_{_\text{PQ}}$ will decay promptly to collimated final states of $2\pi$, $\eta\pi$, $K\bar K$, $3\pi$, $\eta\pi\pi$, etc, which may be tagged as jets, hadronic taus, photon-jets \cite{Toro:2012sv,Ellis:2012sd,Ellis:2012zp} (for final states of 2 or 3 $\pi^{0\,\prime}$s), or even single photons (if the detector's granularity is poor enough, which might be the case for LEP detectors). The much lighter and longer lived MeV axion $a$, with $c\tau_a$ ranging from 0.1 $\mu$m  to 0.1 mm, will produce a highly collimated $e^+e^-$ pair, with a decay vertex displaced by a few mm to several cm. Here, too, the sensitivity of any particular analysis will depend on whether $a$ is tagged as a converted photon, a prompt or displaced lepton-jet \cite{ArkaniHamed:2008qp,Cheung:2009su,Katz:2009qq,Bai:2009it,Baumgart:2009tn,Chan:2011aa,Han:2007ae,Falkowski:2010gv,Curtin:2013fra,Gupta:2015lfa,Dasgupta:2016wxw,Tsai:2016lfg}, or whether it will fail quality criteria for standard objects and simply be vetoed. A thorough study of these possibilities is deferred to \cite{future}\footnote{See also \cite{Bauer:2017nlg,Bauer:2017ris} for exotic $Z$ and Higgs decays to light ALPs.}. It suffices to say, nonetheless, that for $\theta^f_{\Phi H}\sim\mathcal{O}(10^{-2})$  all existing bounds are satisfied for the processes listed in Table~\ref{RareDecayTable}, regardless of assumptions on boosted-PQ-object-tagging.

If $H_e$ in (\ref{PQEW}) is independent of $H_{u,d}$, additional leptonic signatures may arise at LEP (depending on the mass of $H_e$) and/or at the LHC (see \cite{future}).

Finally, note that in order to write the flavor diagonal couplings of $H_{u,d}$ to first generation quarks, we implicitly assumed an MFV-type mechanism which generates the CKM flavor structure of the weak interactions of quarks without spoiling the flavor aligment of Yukawa couplings. Concrete realizations of such mechanism are possible, but are beyond the scope of this study.

\section{\label{Discussion} Discussion}

A short-lived, {\it pion-phobic} QCD axion with mass of several MeV might still offer a viable solution to the strong CP problem. Constraints that have excluded generic MeV axions can be evaded by coupling the axion exclusively to the first generation of SM fermions. Bounds from $K^+\rightarrow\pi^+a$, previously believed to be severe, in fact suffer from large hadronic uncertainties and are currently sufficiently ambiguous to experimentally allow portions of the axion parameter space.

 The extreme {\it pion-phobia} needed to avoid exclusion is a realistic possibility in models with a special relation between the light quark masses and PQ charges, namely,
 \be
 \frac{m_u}{m_d}~\simeq~\frac{\Qpq_d}{\Qpq_u}\;.\nonumber
 \ee
In this study we have imposed this relation {\it ad hoc}, but it is easy to envision how it might arise dynamically. For instance, in supersymmetric models of flavor, quartics are often proportional to charges squared, and flavon VEVs (and thus fermions Yukawas) would then naturally be inversely proportional to the charges.

The associated phenomenology of these variants is rich and testable. Several axion signatures are similar to the those of visibly decaying dark photons, and can be searched for by ongoing and near-future dark photon experiments. They also offer alternative explanations to a few discrepancies in data usually attributed to dark photons, such as the Beryllium-8 and KTeV ($\pi^0\rightarrow e^+e^-$) anomalies. At the GeV scale, these models predict new states coupled to light hadrons awaiting to be uncovered. Those, along with the MeV axion, may appear in rare decays of the SM Higgs, $Z^0$ and other BSM states, yielding exotic signatures with thin jets (i.e., ``$\tau_\text{\,h}$\,-\,like''), prompt or displaced lepton jets, and photon jets.

\begin{acknowledgments}

We thank these colleagues for helpful discussions on various aspects of this study: Wolfgang Altmannshofer, Vincenzo Cirigliano, Kyle Cranmer, Lance Dixon, Bertrand Echenart, Daniel Ega\~na, Glennys Farrar, Alex Friedland, Greg Gabadadze, Michael Graesser, Andy Haas, Eva Halkiadakis, Yonit Hochberg, Anson Hook, Yoni Kahn, Attila Krasnahorkey, Eric Kuflik, Anthony Palladino, Duccio Pappadopulo, Maxim Pospelov, Josh Ruderman, Philip Schuster, Tim Tait, Flip Tanedo, Natalia Toro, Chris Tully, Ken Van Tilberg, and Graziano Venanzoni.

N.W. and D.S.M.A. were supported by the NSF under grants PHY-0947827 and PHY-1316753. At earlier stages of this work, D.S.M.A. was also supported supported by NSF-PHY-0969510 (the LHC Theory Initiative). D.S.M.A. is currently supported by LANL's Early Career LDRD program. D.S.M.A. thanks the Aspen Center for Physics (NSF Grant No. PHYS-1066293) for hospitality while progress was made on this study.

\end{acknowledgments}


\renewcommand\theequation{\thesection.\arabic{equation}}


\bibliographystyle{JHEP}
\bibliography{REFS}

\end{document}